\documentclass[aps,prd,superscriptaddress,showpacs,nofootinbib]{revtex4}
\usepackage{graphicx,epsfig,wrapfig,color}
\newcommand{\be}{\begin{equation}}
\newcommand{\ee}{\end{equation}}
\newcommand{\ba}{\begin{eqnarray}}
\newcommand{\ea}{\end{eqnarray}}
\newcommand{\di}{\!{\rm d}}
\newcommand{\la}{\langle}
\newcommand{\ra}{\rangle}
\newcommand{\fslash}[1] {{\not\! #1\,}}
\newcommand{\bDelta}{ {\bf\Delta}}
\newcommand{\fracS}[2]{{\textstyle\frac{#1}{#2}}}
\newcommand{\bgam}{   {{\mbox{\boldmath$\gamma$} }}}
\newcommand{\krig}[1]{ \;\stackrel{\circ}{#1}} 
\newcommand{\doublesum}[2]
    {\!\sum_{{{\scriptscriptstyle {#1}}\atop{\scriptscriptstyle {#2}}}}}

\newcommand{\triplelim}[3]{{\lim_
{{{\scriptscriptstyle {#1}}\atop{\scriptscriptstyle {#2}}}
    \atop{\scriptscriptstyle {#3}}}}}

\begin{document}
\newcommand*{\Bochum}{
    Institut f\"ur Theoretische Physik II, Ruhr-Universit\"at Bochum,
    D-44789 Bochum, Germany}\affiliation{\Bochum}
\newcommand*{\StPetersburg}{Petersburg Nuclear Physics Institute, Gatchina,
        St.~Petersburg 188350, Russia}\affiliation{\StPetersburg}
\newcommand*{\Coimbra}{Departamento de F\'isica and Centro de F\'isica
    Computacional, Universidade de Coimbra, P-3000 Coimbra,
    Portugal}\affiliation{\Coimbra}
\newcommand*{\Porto}{Faculdade de Engenharia da Universidade do Porto,
    P-4000 Porto, Portugal}\affiliation{\Porto}
\title{
    Nucleon form-factors of the energy momentum tensor\\
    in the chiral quark-soliton model}
\author{K.~Goeke}\affiliation{\Bochum}
\author{J.~Grabis}\affiliation{\Bochum}
\author{J.~Ossmann}\affiliation{\Bochum}
\author{M.~V.~Polyakov}\affiliation{\Bochum}\affiliation{\StPetersburg}
\author{P.~Schweitzer}\affiliation{\Bochum}
\author{A.~Silva}\affiliation{\Coimbra}\affiliation{\Porto}
\author{D.~Urbano}\affiliation{\Porto}\affiliation{\Coimbra}
\date{December, 2006}
\begin{abstract}
The nucleon form factors of the energy-momentum tensor are studied in the
large-$N_c$ limit in the framework of the chiral quark-soliton model.
\end{abstract}
\maketitle

\vspace{-7.4cm}
\begin{flushright}
  preprint RUB-TP2-05-2006
\end{flushright}
\vspace{6cm}



\section{Introduction}
\label{Sec-1:introduction}

The nucleon form factors of the energy momentum tensor (EMT)
\cite{Pagels} were subject to modest interest in literature for a
long time -- probably because the only known process, where they
(in principle) could directly be ``measured'', is elastic
scattering of gravitons off the nucleon. The situation changed,
however, with the advent of generalized parton distribution
functions (GPDs)
\cite{Muller:1998fv,Ji:1996ek,Radyushkin:1997ki,Collins:1996fb}
accessible in hard exclusive reactions
\cite{Saull:1999kt,Adloff:2001cn,Airapetian:2001yk,Stepanyan:2001sm,Ellinghaus:2002bq,Chekanov:2003ya,Aktas:2005ty,Hall-A:2006hx},
see
\cite{Ji:1998pc,Radyushkin:2000uy,Goeke:2001tz,Diehl:2003ny,Belitsky:2005qn}
for reviews.
The form factors of the quark part of the EMT of QCD
--- we use the notation $M_2^Q(t)$, $J^Q(t)$ and $d_1^Q(t)$,
see the definition below in Eq.~(\ref{Eq:ff-of-EMT}) --- appear as
certain Mellin moments of the unpolarized quark GPDs.

The form factor $M_2^Q(0)$ is known at zero-momentum
transfer from inclusive deeply inelastic scattering experiments,
telling us that quarks carry only about half of the momentum of
(a very fast moving) nucleon, and that the rest is carried by gluons.
The appealing perspective is to access by means of GPDs information on
$J^Q(t)$, which --- after extrapolating to zero momentum transfer $t=0$
--- would reveal how much of the nucleon spin is due to quarks
\cite{Ji:1996ek}.
The third form factor, $d_1(t)$, is equally interesting --- promising
to provide information on the distribution of strong forces in the nucleon
\cite{Polyakov:2002yz,Polyakov:2002wz} similarly as the electromagnetic
form factors contain information about the electric charge distribution
\cite{Sachs}.
The information content encoded in GPDs is, however, far reacher than
that, see Refs.~\cite{Burkardt:2000za,Ralston:2001xs,Diehl:2002he}.

In this work we study the form factors of the EMT in the framework of the
chiral quark soliton model (CQSM) \cite{Diakonov:yh,Diakonov:1987ty}.
The model provides a field theoretic description of the nucleon
in the limit of a large number of colours $N_c$, where the nucleon
appears as a chiral soliton of a static background pion field
\cite{Witten:1979kh}.
Numerous nucleonic properties, among others form factors
\cite{Christov:1995hr,Christov:1995vm,Schuren:1991sc,Schweitzer:2003sb}
usual quark and antiquark distribution functions
\cite{Diakonov:1996sr,Diakonov:1997vc,Pobylitsa:1998tk,Wakamatsu:1998rx,Goeke:2000wv,Schweitzer:2001sr}
and GPDs \cite{Petrov:1998kf,Penttinen:1999th,Schweitzer:2002nm,Schweitzer:2003ms,Ossmann:2004bp,Wakamatsu:2005vk},
have been described in this model without adjustable parameters.
As far as those quantities are known an agreement with phenomenology
was observed typically to within an accuracy of $(10-30)\%$.

Our study provides several new results.
In particular, we compute the spatial density distributions, and mean
square radii of the operators of different components of the energy
momentum tensor and its trace. This provides insights not only on, for
example, how the ``mass'' or the ``angular momentum'' are distributed
in the nucleon.
Of particular interest are the results for the spatial distribution
of strong forces in the nucleon. As a byproduct we learn how the soliton
acquires stability in the CQSM. We also observe a physically appealing
connection between the criterion for the stability of the nucleon,
and the sign of the form factor $d_1(t)$ at zero-momentum transfer.

We present results for the form factors which are of practical interest
especially in the case of $J^Q(t)$. As exclusive reactions yield information on
$J^Q(t)$ only at finite $t<0$, some guidance from reliable model calculation
might be of interest for the extrapolation $t\to 0$ required to conclude
how much quarks contribute to the nucleon spin.

We explore the chiral character of the model to study chiral properties
of the form factors. In particular, we derive the leading non-analytic
chiral contributions to the form factors in the large $N_c$ limit.
These non-analytic (in the current quark mass) terms are
model-independent. In fact, our results coincide with results from chiral
perturbation theory \cite{Chen:2001pv,Belitsky:2002jp,Diehl:2006ya},
provided one takes into account that the latter is formulated for finite
$N_c=3$ \cite{Dashen:1993jt,Cohen:1992uy}.

The implicit pion mass dependence of the form factors is of
interest in the context of the chiral extrapolation of lattice QCD data
\cite{Mathur:1999uf,Gadiyak:2001fe,Hagler:2003jd,Gockeler:2003jf,Negele:2004iu,Schroers:2007qf}.
This topic can be addressed in the CQSM \cite{Goeke:2005fs} which
will be done in a separate work \cite{accompanying-paper}.

The note is organized as follows.
Sec.~\ref{Sec-2:FF-of-EMT-in-general}
provides a general discussion of the EMT form factors.
Sec.~\ref{Sec-3:model} introduces the model. In
Sec.~\ref{Sec-4:EMT-in-model} we derive the model expressions for the form factors,
and discuss the numerical results for the densities of the static EMT in
Secs.~\ref{Sec-5:energy-density},~\ref{Sec-6:spin-density},~\ref{Sec-7:pressure+shear}.
In Sec.~\ref{Sec-8:form-factors} we present the results for the form factors,
and conclude our findings in Sec.~\ref{Sec-9:conclusions}.
The Appendices contain: a digression on alternative notations,
a discussion of general properties of the densities of the static EMT,
technical details on the model expressions, and explicit proofs for the
consistency of the model.

\newpage
\section{Form factors of the energy-momentum tensor}
\label{Sec-2:FF-of-EMT-in-general}

The nucleon matrix element of the symmetric energy-momentum tensor
of QCD is characterized by three scalar form factors
\cite{Pagels,Ji:1996ek}. The nucleon matrix elements of the quark
and gluon parts of the symmetric QCD energy-momentum tensor (EMT)
can be parameterized as \cite{Ji:1996ek,Polyakov:2002yz}
(see App.~\ref{App:Alternative-definition} for an alternative notation)
\ba
    \la p^\prime| \hat T_{\mu\nu}^{Q,G}(0) |p\rangle
    &=& \bar u(p^\prime)\biggl[M_2^{Q,G}(t)\,\frac{P_\mu P_\nu}{M_N}+
    J^{Q,G}(t)\ \frac{i(P_{\mu}\sigma_{\nu\rho}+P_{\nu}\sigma_{\mu\rho})
    \Delta^\rho}{2M_N}
    \nonumber \\
    &+& d^{Q,G}_1(t)\,
    \frac{\Delta_\mu\Delta_\nu-g_{\mu\nu}\Delta^2}{5M_N}
    \pm \bar c(t)g_{\mu\nu} \biggr]u(p)\, .
    \label{Eq:ff-of-EMT} \ea
Here $\hat T_{\mu\nu}^Q$ ($\hat T_{\mu\nu}^G$) is the quark (gluon) part of
the QCD energy-momentum tensor. The nucleon states and spinors are normalized
by $\la p^\prime|p\ra = 2p^0(2\pi)^3\delta^{(3)}({\bf p}^\prime-{\bf p})$ and
$\bar u(p) u(p)=2 M_N$ where we suppress spin indices for brevity.
The kinematical variables are defined as $P=(p+p')/2$, $\Delta=(p'-p)$,
$t=\Delta^2$.
The form factor $\bar c(t)$ accounts for non-conservation of the separate
quark and gluon parts of the EMT, and enters the quark and gluon parts
with opposite signs such that the total (quark+gluon) EMT is conserved.

The nucleon form factors of the EMT are related to the unpolarized GPDs
$H^f(x,\xi,t)$ and $E^f(x,\xi,t)$, which are defined as
\ba\label{Def:H-and-E} &&  \hspace{-1cm}
    \int\!\frac{\di\lambda}{2\pi}\,e^{i\lambda x}\la {\bf p'},s'|
    \bar{\psi}_q(-\fracS{\lambda n}{2}) \fslash{n}
    [-\fracS{\lambda n}{2},\fracS{\lambda n}{2}]
    \psi_q(\fracS{\lambda n}{2})|{\bf p},s\ra \nonumber\\
&&  =  H^q(x,\xi,t)\;\bar{u}({\bf p'},s')\fslash{n}u({\bf p},s)
    \nonumber\\
&&  +  E^q(x,\xi,t)\;
    \bar{u}({\bf p'},s')\,\frac{i\sigma^{\mu\nu}n_\mu\Delta_\nu}{2M_N}\,
    u({\bf p},s) , \ea
where $[z_1,z_2]$ denotes the gauge-link,
and the renormalization scale dependence is not indicated for brevity.
The light-like vector $n^\mu$ satisfies $n(p'+p) = 2$, and the skewedness
parameter $\xi$ is defined as $n\Delta = -2\xi$.
To be specific, the form factors in Eq.~(\ref{Eq:ff-of-EMT}) are related to the second
Mellin moments of the unpolarized GPDs in (\ref{Def:H-and-E}) through \cite{Ji:1996ek}
\ba
&&  \int\limits_{-1}^1 \!dx\; x\, \sum\limits_f H^f(x,\xi,t)
    = M_2^Q(t) +\frac 45\, d_1^Q(t)\, \xi^2
    \label{Eq:EMT-and-GPD-H} \\
&&  \int\limits_{-1}^1 \!dx\; x\, \sum\limits_f E^f(x,\xi,t)
    = 2J^Q(t)-M_2^Q(t) - \frac 45\, d_1^Q(t)\, \xi^2\, .
    \label{Eq:EMT-and-GPD-E} \ea
Adding up Eqs.~(\ref{Eq:EMT-and-GPD-H}) and (\ref{Eq:EMT-and-GPD-E}) one
recovers the spin sum rule \cite{Ji:1996ek} promising to access $J^Q(0)$, i.e.\
the total (spin+orbital angular momentum) contribution of quarks to the nucleon
spin,  through the extraction of GPDs from hard exclusive processes and
extrapolation to the unphysical point $t=0$.
The sensitivity of different observables to the total angular momentum of in
particular the $u$-flavour were exposed in the model studies
\cite{Goeke:2001tz,Ellinghaus:2005uc}.
For gluons there are definitions and expressions analog to
(\ref{Def:H-and-E},~\ref{Eq:EMT-and-GPD-H},~\ref{Eq:EMT-and-GPD-E}).
Eqs.~(\ref{Eq:EMT-and-GPD-H},~\ref{Eq:EMT-and-GPD-E}) are
special cases of the so-called polynomiality property of GPDs \cite{Ji:1998pc}
stating that the $N$th Mellin moments of GPDs are polynomials in even powers of
$\xi$ of degree less or equal to $N$:
\ba
    \int\limits_{-1}^1\!\di x\:x^{N-1}\,H(x,\xi,t)
    = h^{(N)}_0(t) + h^{(N)}_2(t)\, \xi^2 + \dots +
        \cases{h^{(N)}_N(t)\, \xi^N    \!\!\!\!\! & for $N$ even\cr
           h^{(N)}_{N-1}(t)\, \xi^{N-1}\!\!\!\!\! & for $N$ odd,} &&
    \label{Def:polynom-Hq}\\
    \int\limits_{-1}^1\!\di x\:x^{N-1}\,E(x,\xi,t)
    = e^{(N)}_0(t) + e^{(N)}_2(t)\, \xi^2 + \dots +
        \cases{e^{(N)}_N(t)\, \xi^N    \!\!\! & for $N$ even\cr
           e^{(N)}_{N-1}(t)\, \xi^{N-1}\!\!\! & for $N$ odd,}&&
    \label{Def:polynom-Eq} \ea
where flavour indices are suppressed for brevity. For a spin $\frac12$ particle
the coefficients in front of the highest power in $\xi$ for even moments $N$
are related to each other and arise from the so-called $D$-term $D^q(z,t)$
with $z=x/\xi$ \cite{Polyakov:1999gs,Teryaev:2001qm},
which has finite support only for $|x|<|\xi|$, according to
\be
    h^{q\,(N)}_N(t)=-\,e^{q\,(N)}_N(t)=\int\limits_{-1}^1\di z\; z^{N-1}D^q(z,t)
        \;\;.\label{Eq:relation-h-e}
\ee

The form factors of the EMT in Eq.~(\ref{Eq:ff-of-EMT}) can be interpreted
\cite{Polyakov:2002yz} in analogy to the electromagnetic form factors
\cite{Sachs} in the Breit frame characterized by $\Delta^0=0$.
In this frame one can define the static energy-momentum tensor for quarks
(and analogously for gluons)
\be\label{Def:static-EMT}
    T_{\mu\nu}^Q({\bf r},{\bf s}) =
    \frac{1}{2E}\int\frac{\di^3\bDelta}{(2\pi)^3}\;\exp(i\bDelta{\bf r})\;
    \la p^\prime,S^\prime|\hat{T}_{\mu\nu}^Q(0)|p,S\ra
\ee
with the initial and final polarization vectors of the nucleon $S$ and
$S^\prime$ defined such that they are equal to $(0,{\bf s})$ in the
respective rest-frame, where the unit vector ${\bf s}$ denotes
the quantization axis for the nucleon spin.

The components of $T_{0 k}^Q({\bf r},{\bf s})$ and
$\varepsilon^{i j k} r_j T_{0k}^Q({\bf r},{\bf s})$ correspond respectively
to the distribution of quark momentum and quark angular momentum inside the
nucleon.
The components of $(T_{ik}^Q-\frac 13\delta_{ik}T^Q_{ll})({\bf r},{\bf s})$
characterize the spatial distribution of ``shear forces" experienced by
quarks inside the nucleon. The respective form factors are related to
$T_{\mu\nu}^Q({\bf r},{\bf s})$ by
\ba
    J^Q(t)+\frac{2t}{3}\, {J^Q}^\prime(t)
    &=& \int\di^3{\bf r}\, e^{-i{\bf r}\bDelta}\,
    \varepsilon^{ijk}\,s_i\,r_j\,T_{0k}^Q({\bf r},{\bf s})\, ,
    \label{Eq:ff-J}\\
    d_1^Q(t)+\frac{4t}{3}\, {d_1^Q}^\prime(t)
    +\frac{4t^2}{15}\, {d_1^Q}^{\prime\prime}(t)
    &=& -\frac{M_N}{2}\, \int\di^3{\bf r}\,e^{-i{\bf r}\bDelta}\,
    T_{ij}^Q({\bf r})\,\left(r^i r^j-\frac{{\bf r}^2}3\,\delta^{ij}\right)\,
    , \label{Eq:ff-d1}\\
    M_2(t)-\frac{t}{4M_N^2}\left(M_2(t)-2 J(t)+\frac 45\, d_1(t) \right)
    &=&\frac{1}{M_N}\,\int\di^3{\bf r}\, e^{-i{\bf r}\bDelta}
    \, T_{00}({\bf r},{\bf s})\,,
    \label{Eq:ff-M2}
\ea
where the prime denotes derivative with respect to the Mandelstam variable $t$.
Note that for a spin-1/2 particle only the $T^{0\mu}$-components are sensitive
to the polarization vector. Note also that Eq.~(\ref{Eq:ff-M2}) holds for the
sum $T_{00}\equiv T_{00}^Q+T_{00}^G$ with $M_2(t)\equiv M_2^Q(t)+M_2^G(t)$ and
$J(t)$ and $d_1(t)$ defined analogously, but not for the separate
quark and gluon contributions -- since otherwise the form factor
$\bar c(t)$ would not cancel out.

The form factor $M_2(t)$ at $t=0$ can be connected to the fractions of the
nucleon momentum carried respectively by quarks and gluons.
This can be seen most conveniently by considering (\ref{Eq:ff-of-EMT}) in the
infinite momentum frame, and one obtains
\be\label{Eq:M2-at-t=0}
    M_2^Q(0) = \int\limits_0^1\!\di x\;
               \sum\limits_q x(f_1^q+f_1^{\bar q})(x)\;,\;\;\;
    M_2^G(0) = \int\limits_0^1\!\di x\;xf_1^g(x)\;,
\ee
where $f_1^a(x)=H^a(x,0,0)$ are the unpolarized parton distributions
accessible in inclusive deeply inelastic scattering.

The form factors $M_2^{Q,G}(t)$, $J^{Q,G}(t)$ and $d_1^{Q,G}(t)$ are
renormalization scale dependent (the indication of the renormalization
scale $\mu$ is suppressed for brevity). Their quark+gluon sums, however, are
scale independent form factors, which at $t=0$ satisfy the constraints,
\ba
    M_2(0)&=&
    \frac{1}{M_N}\,\int\di^3{\bf r}\;T_{00}({\bf r},{\bf s})=1
    \;,\nonumber\\
    J(0)&=&
    \int\di^3{\bf r}\;\varepsilon^{ijk}\,s_i\,r_j\,
    T_{0k}({\bf r},{\bf s})=\frac12\, , \nonumber\\
    d_1(0)&=&
    -\frac{M_N}{2}\, \int\di^3{\bf r}\;T_{ij}({\bf r})\,
    \left(r^i r^j-\frac{{\bf r}^2}3\,\delta^{ij}\right)\equiv d_1\,
    \label{Eq:M2-J-d1}
\ea
which mean that in the rest frame the total energy of the nucleon is equal
to its mass, and that the spin of the nucleon is 1/2. The value of $d_1$
is not known a priori and must be determined experimentally. However,
being a conserved quantity it is to be considered on the same footing as
other basic nucleon properties like mass, anomalous magnetic moment, etc.
Remarkably, $d_1$ determines the behaviour of the $D$-term (and thus the
unpolarized GPDs) in the asymptotic limit of renormalization scale
$\mu\to\infty$ \cite{Goeke:2001tz}.

The form factor $d_1(t)$ is connected to the distribution of pressure and
shear forces experienced by the partons in the nucleon \cite{Polyakov:2002yz}
which becomes apparent by recalling that $T_{ij}({\bf r})$ is the static
stress tensor which (for spin 0 and 1/2 particles) can be decomposed as
\be\label{Eq:T_ij-pressure-and-shear}
    T_{ij}({\bf r})
    = s(r)\left(\frac{r_ir_j}{r^2}-\frac 13\,\delta_{ij}\right)
        + p(r)\,\delta_{ij}\, . \ee
The functions $p(r)$ and $s(r)$ are related to each other due to the conservation
of the total energy-momentum tensor by the differential equation
\be\label{Eq:p(r)+s(r)}
    \frac23\;\frac{\partial s(r)}{\partial r\;}+
    \frac{2s(r)}{r} + \frac{\partial p(r)}{\partial r\;} = 0\;.
\ee
Hereby $p(r)$ describes the radial distribution of the ``pressure" inside the hadron,
while $s(r)$ is related to the distribution of the ``shear forces''
\cite{Polyakov:2002yz}. Another important property which can be directly derived
from the conservation of the EMT is the so-called stability condition. Integrating
$\int\di^3{\bf r}\,r^k(\nabla_iT^{ij})\equiv 0$ by parts one finds that the
pressure $p(r)$ must satisfy the relation
\be\label{Eq:stability}
    \int\limits_0^\infty \!\di r\;r^2p(r)=0 \;.
\ee
Further worthwhile noticing properties which follow from the conservation of
the EMT are discussed in App.~\ref{App:general-relations-from-EMT-conservation}.
Here we only mention that one can express $d_1(t)$ in terms of $p(r)$
and $s(r)$ as (notice the misprint in Eq.~(18) of \cite{Polyakov:2002wz})
\ba\label{Eq:d1-from-s(r)-and-p(r)}
    d_1(t) &=&  5M_N\int\di^3{\bf r}\;\frac{j_2(r\sqrt{-t})}{t}\; s(r)
            =  15M_N\int\di^3{\bf r}\;\frac{j_0(r\sqrt{-t})}{2t}\; p(r)\;,
           \nonumber\\
        d_1 &=& -\,\frac{1}{3}\;M_N \int\di^3{\bf r}\;r^2\, s(r)
         =     \frac{5}{4}\;M_N \int\di^3{\bf r}\;r^2\, p(r)\;.
\ea

Let us review briefly what is known about $d_1$.
For the pion $d_{1,\pi}$ can be calculated exactly using soft pion theorems,
and one obtains $d_{1,\pi}^Q=-M_{2,\pi}^Q$ \cite{Polyakov:1999gs}.
For the nucleon the large-$N_c$ limit predicts \cite{Goeke:2001tz}
\be\label{Eq:D-term-in-large-Nc}
    |d_1^u+d_1^d|={\cal O}(N_c^2) \;\;\; \gg \;\;\;
    |d_1^u-d_1^d|={\cal O}(N_c)
\ee
which is in agreement with lattice QCD
\cite{Hagler:2003jd,Gockeler:2003jf,Negele:2004iu}.
The constant $d_1^Q=d_1^u+d_1^d$ is found negative on the lattice
\cite{Hagler:2003jd,Gockeler:2003jf,Negele:2004iu}.

From model calculations in the CQSM it was estimated that $d_1^Q\approx -4.0$
at scales of few ${\rm GeV}^2$ \cite{Petrov:1998kf,Kivel:2000fg}.
n a simple ``liquid drop'' model $d_1$ is related to the surface tension
of the ``liquid'' and comes out negative \cite{Polyakov:2002yz}.
Such a model is in particular applicable to large nuclei. Predictions for
the behaviour of the cross section of deeply virtual Compton scattering off
nuclei made on the basis of this model \cite{Polyakov:2002yz}, have been
confirmed in practical calculations assuming realistic models for nuclei
\cite{Guzey:2005ba}. In particular, also the $D$-terms of nuclei were
found negative \cite{Guzey:2005ba}.

Finally, let us discuss an interesting connection of the constant $d_1$
and the mean square radius $\la r_F^2\ra$ of the trace of the total EMT
operator. Due to the trace anomaly
\cite{Adler:1976zt,Nielsen:1977sy,Collins:1976yq,Adler:2004qt}
the latter is given by
\be\label{Eq:ff-of-EMT-trace-1}
    \hat{T}_{\!\mu}^{\,\mu} \equiv \frac{\beta}{2g}\;F^{\mu\nu}F_{\mu\nu}
    +(1+\gamma_m)\sum_am_a\bar\psi_a\psi_a \;.\ee
For notational simplicity let us introduce the scalar form factor $F(t)$
\be\label{Eq:ff-of-EMT-trace-0}
        \la p^\prime|\hat{T}_{\!\mu}^{\,\mu}(0) |p\rangle
        = M_N \,\bar u(p^\prime)u(p)\;F(t)\;.
\ee
which can be expressed in terms of the form factors in (\ref{Eq:ff-of-EMT}) as
\be\label{Eq:ff-of-EMT-trace-2}
    F(t) = M_2(t)+\frac{t}{4M_N^2}\biggl(2J(t)-M_2(t)\biggr)
    -\frac{3t}{5M_N^2}\;d_1(t)\;.
\ee
It satisfies $F(0)=1$ and its derivative at $t=0$ defines the mean
square radius of the EMT trace operator
\be\label{Eq:ff-of-EMT-trace-3}
    \la r_F^2\ra = 6F^\prime(0) =
    6\biggl(M_2^\prime(0)-\frac{3\,d_1}{5M_N^2}\biggr)\;.
\ee
Analogously, one may define the mean square radius $\la r_E^2\ra$ of the energy
density operator $\hat{T}^{00}$ for which one finds from
Eq.~(\ref{Eq:ff-M2}) the following result
\be\label{Eq:ff-of-EMT-trace-4}
    \la r_E^2\ra=6\biggl(M_2^\prime(0)-\frac{d_1}{5M_N^2}\biggr)\;.
\ee
Exploring (\ref{Eq:ff-of-EMT-trace-4}) we see that $\la r_F^2\ra$
is related to the mean square radius of the energy density
$\la r_E^2\ra$ as follows
\be\label{Eq:ff-of-EMT-trace-5}
    \la r_F^2\ra = \la r_E^2\ra -\frac{12\,d_1}{5M_N^2}\;.
\ee
Since $d_1$ is observed to be negative, one has
$\la r_F^2\ra > \la r_E^2\ra$.

\newpage
\section{The nucleon as a chiral soliton}
\label{Sec-3:model}

The effective theory underlying the CQSM was derived from the
instanton model of the QCD vacuum \cite{Diakonov:1983hh,Diakonov:1985eg}
which assumes that the basic properties of the QCD vacuum are dominated
by a strongly interacting medium of instantons and anti-instantons.
This medium is diluted with a density proportional to $(\rho_{\rm av}/R_{\rm av})^4$
where $\rho_{\rm av}\approx 0.3\,{\rm fm}$ is the average instanton size and
$R_{\rm av}$ the average instanton separation.
It is found $\rho_{\rm av}/R_{\rm av}\sim \frac13$
\cite{Diakonov:1983hh,Diakonov:1985eg,Diakonov:1995qy},
see \cite{Diakonov:2000pa} for reviews.

Due to interactions with instantons in this medium light quarks acquire a
dynamical (``constituent'') quark mass which is strictly speaking momentum-dependent,
i.~e.\  $M=M(p)$, and drops to zero for momenta $p\gg \rho_{\rm av}^{-1}$.
At low momenta below a scale set by $\rho_{\rm av}^{-1}\approx 600\,{\rm MeV}$
the dynamics of these effective quark degrees of freedom is governed by the partition
function \cite{Diakonov:1984tw,Dhar:1985gh}
\be\label{eff-theory}
    Z_{\rm eff} = \int\!\!{\cal D}\psi\,{\cal D}\bar{\psi}\,
    {\cal D}U\,\exp\biggl(iS_{\rm eff}(\bar\psi,\psi,U)\biggr)\;,\;\;\;
    S_{\rm eff}(\bar\psi,\psi,U)=\int\di^4x\;\bar{\psi}\,
    (i\fslash{\partial}-M\,U^{\gamma_5}-m)\psi\,.\ee
Here we restrict ourselves to two light flavours, $U=\exp(i\tau^a\pi^a)$ denotes
the chiral pion field with $U^{\gamma_5} = \exp(i\gamma_5\tau^a\pi^a)$,
and $m=m_u=m_d$ is the current quark mass neglecting isospin breaking effects.
The smallness of the instanton packing fraction $\rho_{\rm av}/R_{\rm av}$ plays an
important role in the derivation of (\ref{eff-theory}) from the instanton vacuum model.

In practical calculations it is convenient to replace $M(p)$ by a constant mass
$M=M(0)=350\,{\rm MeV}$ following from the instanton vacuum \cite{Diakonov:2000pa},
and to regularize the effective theory by means of an explicit (e.~g.\ proper-time,
or Pauli-Villars) regularization with a cutoff of ${\cal O}(\rho_{\rm av}^{-1})$
whose precise value is fixed to reproduce the physical value of the
pion decay constant $f_\pi=93\,{\rm MeV}$ given by a logarithmically UV-divergent
expression in the effective theory (\ref{eff-theory}).
For most quantities the effects of different regularizations are of
${\cal O}(M^2\rho_{\rm av}^2)\propto\rho_{\rm av}^4/R_{\rm av}^4$, i.e.\
parametrically small.

The CQSM is an application of the effective theory (\ref{eff-theory})
to the description of baryons \cite{Diakonov:yh,Diakonov:1987ty}.
The Gaussian integral over fermion fields in (\ref{eff-theory})
can be solved exactly. The path integral over pion field configurations,
however, can be solved only by means of the saddle-point approximation
(in the Euclidean formulation of the theory).
This step is strictly justified in the large-$N_c$ limit.
In the leading order of the large-$N_c$ limit the pion field is static,
and one can determine the spectrum of the one-particle Hamiltonian of the
effective theory (\ref{eff-theory})
\be\label{Hamiltonian}
    \hat{H}|n\ra=E_n |n\ra \;,\;\;
    \hat{H}=-i\gamma^0\gamma^k\partial_k+\gamma^0MU^{\gamma_5}+\gamma^0m
    \;. \ee
The spectrum consists of an upper and a lower Dirac continuum, distorted by
the pion field as compared to continua of the free Dirac-Hamiltonian $\hat{H}_0$
(which follows from $\hat{H}$ in (\ref{Hamiltonian}) by replacing $U^{\gamma_5}\to 1$)
and of a discrete bound state level of energy $E_{\rm lev}$,
if the pion field is strong enough.
By occupying the discrete level and the states of the lower continuum each
by $N_c$ quarks in an anti-symmetric colour state, one obtains a state
with unity baryon number.
The soliton energy $E_{\rm sol}$ is a functional of the pion field
\be\label{Eq:soliton-energy}
    E_{\rm sol}[U] = N_c \biggl[E_{\rm lev}+
    \sum\limits_{E_n<0}(E_n-E_{n_0})\biggr]_{\rm reg} \;.
    \ee
$E_{\rm sol}[U]$ is logarithmically divergent, see
App.~X for the explicit expression
in the proper-time regularization.
By minimizing $E_{\rm sol}[U]$ one obtains the self-consistent solitonic
pion field $U_c$.  This procedure is performed for symmetry reasons in
the so-called hedgehog ansatz
\be\label{hedgehog}
    \pi^a({\bf x})=e^a_r\;P(r) \;,\;\;
    U({\bf x})=\cos P(r)+i \tau^a e_r^a \sin P(r)\;,\ee
with the radial (soliton profile) function $P(r)$ and $r=|{\bf x}|$,
${\bf e}_r = {\bf x}/r$.
The nucleon mass $M_N$ is given by $E_{\rm sol}[U_c]$.
The self-consistent profile satisfies $P_c(0)=-\pi$ and behaves as
\be\label{Eq:profile-at-large-r}
    P(r) = -\,\frac{A}{r^2}\,(1+m_\pi r)\exp(-m_\pi r)
    \;\;\;\mbox{at large $r$, \hspace{0.2cm} with}\;\;\;
    A = \frac{3g_A}{8\pi f_\pi^2}
\ee
where $g_A=1.26$ is the axial coupling constant and the pion mass $m_\pi$ is
connected to $m$ in (\ref{eff-theory}) by the Gell-Mann--Oakes--Renner relation
for small $m$.
In the large-$N_c$ limit the path integral over $U$ in Eq.~(\ref{eff-theory}) is
solved by evaluating the expression at $U_c$ and integrating over translational
and rotational zero modes of the soliton solution in the path integral.
In order to include corrections in the $1/N_c$-expansion one considers time dependent
pion field fluctuations around the solitonic solution. In practice hereby one
restricts oneself to time dependent rotations of the soliton field in spin- and
flavour-space which are slow due to the large moment of inertia of the soliton,
$I={\cal O}(N_c)$, given by
\be\label{Eq:mom-inertia}
    I=\frac{N_c}{6}\doublesum{m,\rm non}{n,\rm occ}
    \frac{\la n|\tau^a|m\ra\,\la m|\tau^a|n\ra}{E_m-E_n}\biggl|_{\rm reg} \;.
\ee
As indicated, $I$ is logarithmically divergent and has to be regularized.
In (\ref{Eq:mom-inertia}) the sum goes over occupied (``occ'') states $n$
which satisfy $E_n\le E_{\rm lev}$, and over non-occupied (``non'') states
$m$ which satisfy $E_m > E_{\rm lev}$.

\section{\boldmath Form factors of the energy momentum tensor in the CQSM}
\label{Sec-4:EMT-in-model}

The gluon part of the EMT is zero in the effective theory (\ref{eff-theory}),
since there are no explicit gluon degrees of freedom. Consequently in the model
the quark energy momentum tensor is conserved by itself, and the form-factor
$\bar c(t)$ in Eq.~(\ref{Eq:ff-of-EMT}) vanishes. This is demonstrated explicitly
in App.~\ref{App:conservation-of-EMT-in-model}. The nucleon matrix elements of the
effective operator for the quark energy momentum tensor (we omit in the following
the index $Q$) is given by the path integral
\be\label{Eq:EMT-path-int}
    \la p^\prime| \hat T_{\mu\nu}(0) |p\rangle
    = \lim\limits_{T\to\infty}\frac{1}{Z_{\rm eff}\!\!}
    \int\!\di^3{\bf x}\;\di^3{\bf y}\;
    e^{i{\bf p}^\prime{\bf y}-i{\bf px}}
    \int\!\!{\cal D}\psi\,{\cal D}\bar{\psi}\,{\cal D}U\,
    J_{N^\prime}(-T/2,{\bf y})\hat{T}_{\mu\nu}^{\rm eff}(0)
    J_{N}^\dag  ( T/2,{\bf x})
    \exp\biggl(iS_{\rm eff}(\bar\psi,\psi,U)\biggr)\;,
\ee
where $J_{N}(x)$ denotes the nucleon current, see
\cite{Diakonov:1987ty,Christov:1995hr,Christov:1995vm} for explicit expressions.
The symmetric energy momentum tensor for quarks in the effective theory
(\ref{eff-theory}) is given by
(the arrows indicate on which fields the derivatives act)
\be
    \hat{T}_{\mu\nu}^{\rm eff}(x) = \frac{1}{4}
    \bar\psi(x)\,\biggl(
     i\gamma^\mu\overrightarrow{\partial}^\nu
    +i\gamma^\nu\overrightarrow{\partial}^\mu
    -i\gamma^\mu\overleftarrow{ \partial}^\nu
    -i\gamma^\nu\overleftarrow{ \partial}^\mu
    \biggr)\,\psi(x)\;.
\ee

For the calculation of the EMT nucleon matrix elements in the model we
have to evaluate consistently the nucleon-bispinor expressions appearing on the
right-hand-side (RHS) of Eq.~(\ref{Eq:ff-of-EMT}) in the large-$N_c$ limit where
$p^0=M_N={\cal O}(N_c)$ and $|p^i|={\cal O}(N_c^0)$ such that $|t|\ll M_N^2$. Keeping
in mind that in the large-$N_c$ limit the form-factors behave as \cite{Goeke:2001tz}
\be\label{Eq:form-factors-in-large-Nc}
    M_2(t) =  {\cal O}(N_c^0)   \;,\;\;\;\;
    J(t)   =  {\cal O}(N_c^0)   \;,\;\;\;\;
    d_1(t) =  {\cal O}(N_c^2)   \;,\;\;\;\;
\ee
we obtain from (\ref{Eq:ff-of-EMT}) the following relations for the form factors
\ba
    \langle p^\prime,S_3^\prime|T^{00}_{\rm eff}|p,S_3\rangle
    &=& 2M_N^2\delta_{S_{3}^\prime S_{3}^{\phantom{\prime}}}
    \biggl(M_{2}(t)-\frac{t}{5M_{N}^{2}}d_{1}(t)\biggr)
    \label{Eq:T00}\\
    \langle p^\prime,S_3^\prime|T^{ik}_{\rm eff}|p,S_3\rangle
    &=& 2\delta_{S_{3}^\prime S_{3}^{\phantom{\prime}}}
    \biggl(\Delta^i\Delta^k-\delta^{ik}{\bf\Delta}^2\biggr)\,d_{1}(t)
    \label{Eq:Tik}\\
    \langle p^\prime,S_3^\prime|T^{0k}_{\rm eff}|p,S_3\rangle
    &=& -i\,M_N\,\varepsilon^{klm}\Delta^{l}
    \sigma_{S_{3}^\prime S_{3}^{\phantom{\prime}}}^{m}\,J(t)\;.
    \phantom{\biggl|}
    \label{Eq:T0k}
\ea
The expressions for $M_2(t)$ and $d_1(t)$ could, of course, be separated which
we shall do more conveniently at a later stage. Evaluating the respective
components of the EMT in (\ref{Eq:EMT-path-int}) yields
(vacuum subtraction analog to (\ref{Eq:soliton-energy}) is implied)
\ba
    \langle p^\prime,S_3^\prime|\hat{T}^{00}_{\rm eff}|p,S_3\rangle
    &=& \delta_{S_3^\prime S_3^{\phantom{\prime}}}\;2M_N N_c
    \sum\limits_{n,\rm occ}E_n\la n|e^{i{\bf\Delta}\hat{\bf x}}|n\ra
    \biggl|_{\rm reg} \label{Eq:T00-in-model}\\
    \langle p^\prime,S_3^\prime|\hat{T}^{ik}_{\rm eff}|p,S_3\rangle
    &=& \delta_{S_3^\prime S_3^{\phantom{\prime}}}\; 2M_N N_c
    \sum\limits_{n,\rm occ}\frac14\,\la n|\biggl(
    \{e^{i{\bf\Delta}\hat{\bf x}},\gamma^0\gamma^i\hat{p}^k\}+
    (i\leftrightarrow k)\biggr)|n\ra\biggl|_{\rm reg}
    \label{Eq:Tik-in-model}\\
    \langle p^\prime,S_3^\prime|\hat{T}^{0k}_{\rm eff}|p,S_3\rangle
    &=& \sigma^{\,l}_{S_3^\prime S_3^{\phantom{\prime}}}
    \frac{M_N N_c}{4I}\doublesum{m,\rm occ}{j,\rm non}
    \frac{\la m|\tau^l|j\ra}{E_m-E_j}
    \la j|\biggl(\{e^{i{\bf\Delta}\hat{\bf x}},\hat{p}^k\}
    +(E_m+E_j)\gamma^0\gamma^ke^{i{\bf\Delta}\hat{\bf x}}\biggr)|m\ra
    \biggl|_{\rm reg}\label{Eq:T0k-in-model} \;.
\ea
These expressions are logarithmically divergent and have to be
regularized appropriately, see App.~\ref{App:regularization} for details.

Noteworthy, the matrix elements for the components $\hat{T}^{00}_{\rm eff}$ and
$\hat{T}^{0k}_{\rm eff}$ related to $M_2(t)$ and $d_1(t)$ are spin-independent
and receive contributions from leading order of the large-$N_c$ expansion.
In contrast to this, in order to address $\hat{T}^{0k}_{\rm eff}$ connected to $J(t)$
one needs matrix elements involving nucleon spin flip which appear only when
considering $1/N_c$ (``rotational'') corrections.
Inserting the results
(\ref{Eq:T00-in-model},~\ref{Eq:Tik-in-model},~\ref{Eq:T0k-in-model})
into Eqs.~(\ref{Eq:T00},~\ref{Eq:Tik},~\ref{Eq:T0k})  yields
\ba
    M_{2}(t)-\frac{t}{5M_{N}^{2}}d_{1}(t)
    &=&
    \frac{N_c}{M_N}\sum\limits_{n,\rm occ}E_{n}
    \la n|e^{i{\bf \Delta}\hat{\bf x}}|n\ra\biggl|_{\rm reg}\;,
    \label{Eq:M2-d1-model}\\
    d_{1}(t)
    &=&
    \frac{5M_N N_c }{4t}
    \sum_{n,\rm occ}\langle n|
    \biggl\{\gamma^0\mbox{\boldmath$\gamma$}\hat{\bf p},
    \,e^{i{\bf \Delta}\hat{\bf x}}\biggr\} |n\rangle\biggl|_{\rm reg} \;,
    \label{Eq:d1-model}\\
    J(t)
    &=&
    \frac{iN_c\,\varepsilon^{klm}\Delta^k}{8I t}
    \doublesum{n,\rm occ}{j,\rm non}
    \frac{\langle n|\tau^l|j\rangle}{E_j-E_n}
    \langle j|\biggl(
    \biggl\{ e^{i{\bf \Delta}\hat{\bf x}},\hat{p}^m\biggr\}
    +(E_n+E_j)e^{i{\bf \Delta}\hat{\bf x}}\gamma^0\gamma^m\biggr)
    |n\rangle\biggl|_{\rm reg}\;.
    \label{Eq:J-model} \ea
The derivation of Eqs.~(\ref{Eq:M2-d1-model},~\ref{Eq:d1-model},~\ref{Eq:J-model})
follows closely the derivation of the model expressions for electromagnetic
\cite{Christov:1995hr} or other form factors, and we omit the details here.
Instead, we demonstrate explicitly in the Appendices
\ref{App:d1(0)-and-d1prime(0)}--\ref{App:FFs-from-EMT-and-GPD-2}
that one obtains the same expressions for the form factors from the model
expressions for GPDs via the sum rules (\ref{Eq:EMT-and-GPD-H}) and
(\ref{Eq:EMT-and-GPD-E}).

We introduce the Fourier transforms of the form factors which are
radial functions (``densities'') defined as
\ba
    \rho_E(r)
    &=&
    N_c\sum\limits_{n,\rm occ}E_n\;
        \phi_n^\ast({\bf r})\phi_n({\bf r})\biggl|_{\rm reg}
    \label{Eq:density-energy}\\
    p(r)
    &=&
    \frac{N_c}{3}\sum_{n,\rm occ}\phi_n^\ast({\bf r})\,(\gamma^0
    \mbox{\boldmath$\gamma$}\hat{\bf p})\,\phi_n({\bf r})\biggl|_{\rm reg}
    \label{Eq:density-pressure}\\
    \rho_J(r)
    &=&
    - \frac{N_c}{24I} \doublesum{n,\rm occ}{j,\rm non}\epsilon^{abc}r^a
    \phi_j^\ast({\bf r})\biggl(2\hat{p}^b+(E_n+E_j)\gamma^0\gamma^b\biggr)
    \phi_n({\bf r})\;\frac{\langle n|\tau^c|j\rangle}{E_j-E_n}\,
    \biggl|_{\rm reg}\;
    \label{Eq:density-spin}
\ea
where it is understood that
$\hat{p}^j\equiv \frac{i}{2}(\overleftarrow{\nabla}^j-\overrightarrow{\nabla}^j)$,
and which allow to reexpress $M_2(t)$, $J(t)$ and $d_1(t)$ in
(\ref{Eq:M2-d1-model},~\ref{Eq:d1-model},~\ref{Eq:J-model}) as
\ba
        M_2(t)-\frac{t}{5M_N^2}d_1(t)
        &=& \frac{1}{M_N}\int\di^3{\bf r}\;\rho_E(r) \;j_0(r\sqrt{-t})
        \label{Eq:M2-d1-model-comp}\\
        d_1(t)
        &=& \frac{15 M_N}{2}\int\di^3{\bf r}\;p(r) \;\frac{j_0(r\sqrt{-t})}{t}
        \label{Eq:d1-model-comp}\\
        J(t)
        &=& 3
    \int\di^3{\bf r}\;\rho_J(r)\;\frac{j_1(r\sqrt{-t})}{r\sqrt{-t}}\;.
    \label{Eq:J-model-comp}
\ea
Here $j_k(z)$ denote Bessel functions with $j_0(z) = \frac{\sin z}{z}$
and $j_1(z)=-j_0^\prime(z)$.

The densities
(\ref{Eq:density-energy},~\ref{Eq:density-pressure},~\ref{Eq:density-spin})
are convenient not only because their numerical evaluation is more economic
than the direct calculation of the form factors
(\ref{Eq:M2-d1-model},~\ref{Eq:d1-model},~\ref{Eq:J-model}).
These densities are interesting objects by themselves, and it is instructive to
discuss their theoretical properties in detail which we shall do in the following.
Simultaneously we will present the numerical results for the densities.

For the numerical calculation we employ the so-called Kahana-Ripka method
\cite{Kahana:1984be}, whose application to calculations in the CQSM is
described in detail e.g.\  in Ref.~\cite{Christov:1995vm}, and use the
proper-time regularization. The latter allows to include effects of symmetry
breaking due to an explicit chiral symmetry breaking current quark mass $m$
in the effective action (\ref{eff-theory}), and to study in the model how
observables vary in the chiral limit.
(For a study --- in the spirit of \cite{Goeke:2005fs} --- of observables at
pion masses as large as they appear in present day lattice calculations
the reader is referred to \cite{accompanying-paper}.)

In order to explore effects of different regularizations
we perform also a calculation with the Pauli-Villars regularization method
which, however, is applicable only in the chiral limit \cite{Kubota:1999hx}.
All results are summarized in Table~\ref{Table:all-results}.

\section{\boldmath Energy density}
\label{Sec-5:energy-density}

The density $\rho_E(r)$ is just the energy density $T^{00}(r)$
in the static energy momentum tensor (\ref{Def:static-EMT}).
By using the ortho-normality of the single-quark states
$\int\di^3{\bf r}\,\phi_n^\ast({\bf r})\phi_{n^\prime}({\bf r})=\delta_{{nn}^\prime}$
and comparing to Eq.~(\ref{Eq:soliton-energy}) we find that
\be\label{Eq:density-energy-1}
    \int\di^3{\bf r}\,\rho_E(r) =
    N_c\sum\limits_{n,\rm occ}E_n\;\int\di^3{\bf r}\,
        \phi_n^\ast({\bf r})\phi_n({\bf r})\biggl|_{\rm reg} = M_N\;.
\ee
The normalization (\ref{Eq:density-energy-1}) ensures the correct constraint of the
form factor $M_2(t)$ at $t=0$. In fact, by taking in Eq.~(\ref{Eq:M2-d1-model-comp})
the limit $t\to 0$
(notice that $d_1(t)$ takes a well-defined finite value for $t\to 0$, see below)
we obtain
\be\label{App-constraints-01}
    M_{2}(0) = \frac{1}{M_N}\int\di^3{\bf r}\;\rho_E(r) = 1\;.
\ee
This is the consistent constraint in the model for $M_2(t)$ at $t=0$,
cf.\  Eq.~(\ref{Eq:M2-at-t=0}), since there are no gluons in the effective theory
such that consequently the entire momentum of the nucleon is carried by quarks
and antiquarks \cite{Diakonov:1996sr}.

\begin{figure*}[t!]
\begin{tabular}{cc}
    \hspace{-1cm}
    \includegraphics[width=7cm]{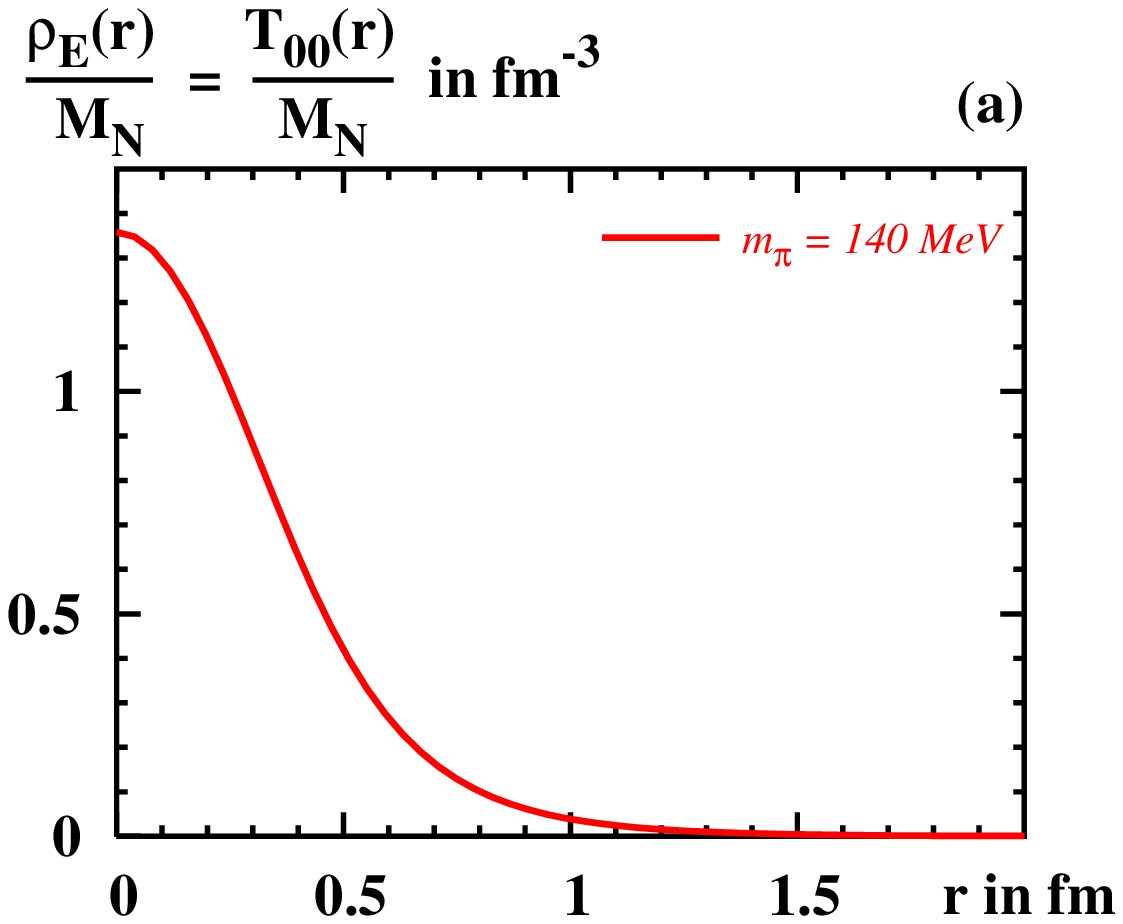}&
    \includegraphics[width=7cm]{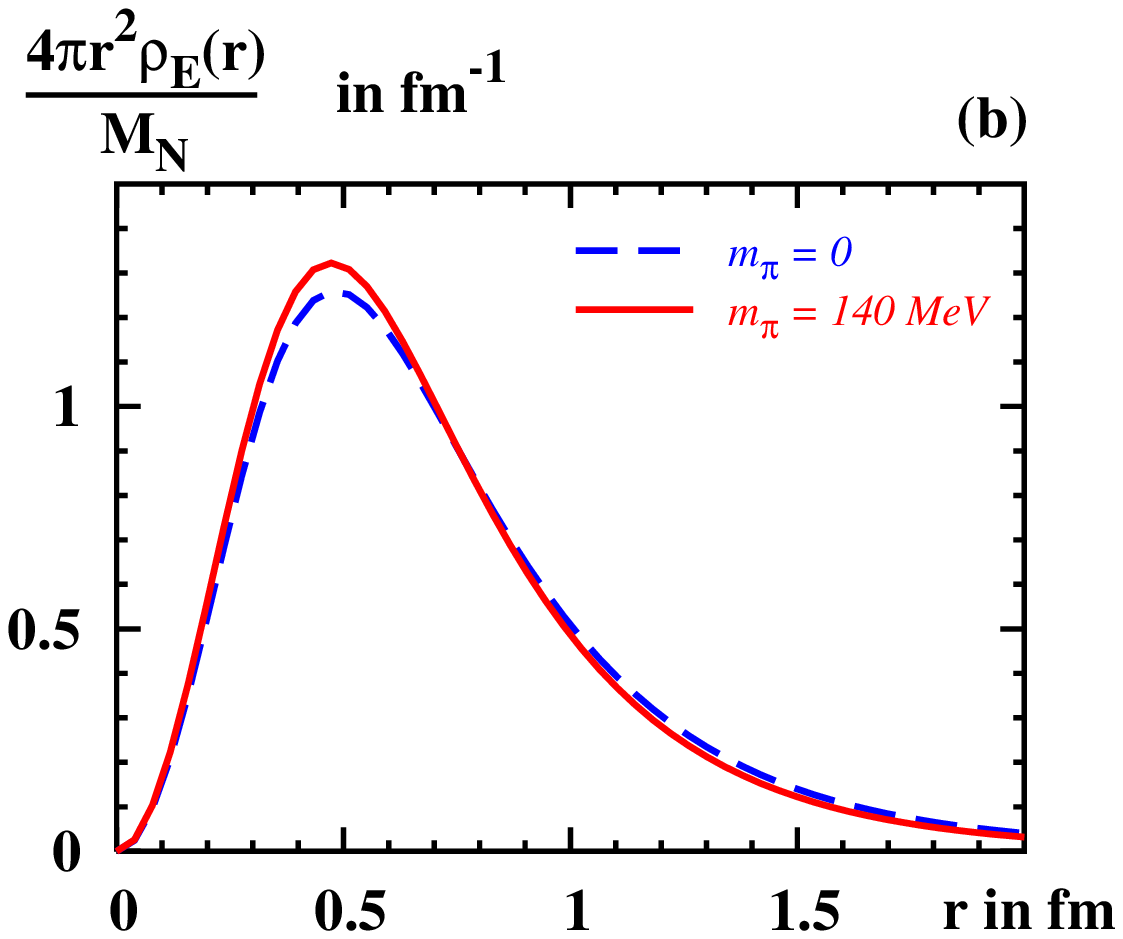}
\end{tabular}
    \caption{\label{Fig1-energy-density}
    \footnotesize\sl
    (a) The normalized energy density $\rho_E(r)/M_N$ of the nucleon
    as function of $r$ for the physical situation with
    $m_\pi=140\,{\rm MeV}$.
    The curve is normalized such that it yields unity upon
    integration over the entire volume.
    (b) The normalized energy density $4\pi r^2\rho_E(r)/M_N$ as function
    of $r$ in the chiral limit and for $m_\pi=140\,{\rm MeV}$.
    The curves are normalized such that one obtains
    unity upon integration over $r$.}
\end{figure*}

Fig.~\ref{Fig1-energy-density}a shows the normalized density $\rho_E(r)/M_N$ as
function of $r$ for the physical situation with a pion mass of $140\,{\rm MeV}$.
In this case in the model the nucleon mass is about $1250\,{\rm MeV}$.
This overestimate of the physical nucleon mass of ${\cal O}(300\,{\rm MeV})$
is typical for the soliton approach and its origin is well-understood
\cite{Pobylitsa:1992bk}.
In the center of the nucleon one finds $\rho_E(0)=1.7\,{\rm GeV\,fm}^{-3}$
or $3.0 \times 10^{15}\,{\rm g\,\,cm}^{-3}$. In order to gain some intuition about
this number we remark that this corresponds roughly to 13 times the equilibrium
density of nuclear matter.

It is instructive to consider the energy density in the chiral limit.
The result is shown in Fig.~\ref{Fig1-energy-density}b where we compare
$4\pi r^2\rho_E(r)/M_N$ as functions of $r$ for $m_\pi = 0$ and $140\,{\rm MeV}$.
The curves are normalized such that one obtains unity when integrating over $r$.
Fig.~\ref{Fig1-energy-density}b shows that with decreasing $m_\pi$
the energy density is spread more widely. This can be quantified by considering
the $m_\pi$-dependence of the mean square radius of the energy density
(\ref{Eq:ff-of-EMT-trace-4}) defined as
\be\label{Eq:def-energy-mean-square-radius}
     \la r_E^2\ra =
    \frac{\int\di^3{\bf r}\;r^2\rho_E(r)}{\int\di^3{\bf r}\;\rho_E(r)}\;,\ee
and which increases in the chiral limit, see Table~\ref{Table:all-results}.
This is an intuitively expected feature. As the pion mass decreases, the range
of the ``pion cloud'' increases and the nucleon becomes ``larger''.

The popular idea of the nucleon consisting of a ``quark core'' surrounded by a
``pion cloud'' is strictly speaking well defined in models only. Here, in the CQSM,
we shall associate the contribution of the discrete level as ``quark core'' and
the contribution of the negative continuum states as ``pion cloud''.
From the long-distance behaviour of the soliton profile (\ref{Eq:profile-at-large-r})
one finds in the chiral limit
\be\label{Eq:energy-density-in-grad-expansion}
    \rho_E(r) = \frac{f_\pi^2}{4}\;
    {\rm tr}_F\,\biggl[\nabla^k U({\bf r})\,\nabla^k U^\dag({\bf r})\biggr]
    +\dots
    \;\stackrel{r\to\rm large}{\longrightarrow}
    \; 3 \left(\frac{3g_A}{8\pi f_\pi}\right)^2 \frac{1}{r^6}
\ee
where ${\rm tr}_F$ is the trace over flavour indices of the SU(2) matrices and the
dots in the intermediate step denote terms which contain higher $U$-field derivative
terms and vanish faster at large $r$ than the displayed leading term.
The result (\ref{Eq:energy-density-in-grad-expansion}) can be read off
from Eq.~(7.8) of \cite{Diakonov:1996sr}.
At $m_\pi\neq 0$ the decay of $\rho_E(r)$ at large $r$ is exponential due to
the corresponding behaviour of the  soliton profile (\ref{Eq:profile-at-large-r}).
This diminishes the ``range of the pion cloud'' and reduces $\la r_E^2\ra$.

These observations can be further quantified by considering the chiral
expansion of (\ref{Eq:def-energy-mean-square-radius}) which gives,
see App.~\ref{App:chrial-properties},
\be\label{Eq:rE2-vs-mpi}
     \la r_E^2\ra = \la \krig{r}{\!}_E^2\ra
    - \frac{81\,g_A^2}{64 \pi f_\pi^2 M_N}\; m_\pi + \dots
\ee
Here and in the following the $^\circ$ above a quantity denotes its value in the
chiral limit, and the dots denote terms vanishing faster in the chiral limit than
the respective leading term. Considering the non-commutativity of the limits
$N_c\to\infty$ and $m_\pi\to 0$
(see the discussion below in Sec.~\ref{Sec-8:form-factors})
Eq.~(\ref{Eq:rE2-vs-mpi}) agrees with chiral perturbation theory
\cite{Belitsky:2002jp}.

The term linear in $m_\pi$ in Eq.~(\ref{Eq:rE2-vs-mpi}),
i.e.\  the leading non-analytic (in the current quark mass $m\propto m_\pi^2$)
contribution to the mean square radius of the energy density accounts almost
entirely for the reduction of $\la r_E^2\ra$ from $m_\pi=0$ to $140\,{\rm MeV}$,
see Table~\ref{Table:all-results}.\footnote{
    In the CQSM the physical value of $g_A$ is underestimated by about $30\%$
    in the proper time regularization in the leading order of the $1/N_c$
    expansion to which we work here. For consistency we have to use here
    this leading order $N_c$ model result for $g_A$.
    Including $1/N_c$ corrections the model describes $g_A$ more accurately
    \cite{Christov:1995vm}.}
At the physical pion mass we find $\la r_E^2\ra = 0.67\,{\rm fm}^2$.
This value is similar to the electric charge radius of the proton.
In fact, we observe a qualitative similarity of the energy density and electric
proton charge distributions in the model \cite{Christov:1995hr,Christov:1995vm}.

\newpage

\section{\boldmath Angular momentum density}
\label{Sec-6:spin-density}

Taking in Eq.~(\ref{Eq:J-model-comp}) the limit $t\to 0$ yields
\be\label{Eq:J-model-comp-01}
    J(0) = \int\di^3{\bf r}\;\rho_J(r)\;,
\ee
which shows in which sense it is adequate to refer to $\rho_J(r)$ as
the ``angular momentum density''. In order to see that $J(0)=\frac12$,
i.e.\ that the constraint (\ref{Eq:M2-J-d1}) is satisfied in the CQSM
we insert (\ref{Eq:density-spin}) in the above equation and obtain
\be\label{Eq:J-model-comp-02}
    2J(0) =
    \frac{N_c}{12I} \doublesum{n,\rm occ}{j,\rm non}\epsilon^{abc}
    \la j|\biggl(2\hat{p}^a\hat{r}^b+(E_n+E_j)\gamma^0\gamma^a\hat{r}^b\biggr)
    |n\ra\;\frac{\langle n|\tau^c|j\rangle}{E_j-E_n}\,\biggl|_{\rm reg}
    \stackrel{\mbox{\footnotesize Ref.~\cite{Ossmann:2004bp}}}{=}
    \int\limits_{-1}^1 \!dx\; x\, \sum\limits_f (H^f+E^f)(x,\xi,0)
    \stackrel{\mbox{\footnotesize Ref.~\cite{Ossmann:2004bp}}}{\equiv} 1 \,.
\ee
In the intermediate step in Eq.~(\ref{Eq:J-model-comp-02}) we recovered the model
expression for the second moment of $\sum_f(H^f+E^f)(x,\xi,t)$ at $t=0$, see
Appendix C of \cite{Ossmann:2004bp}.
This sum rule, which follows from adding up Eqs.~(\ref{Eq:EMT-and-GPD-H}) and
(\ref{Eq:EMT-and-GPD-E}), was explicitly proven to be satisfied in the model in
\cite{Ossmann:2004bp}.
In the model the entire nucleon spin is due to the spin and orbital angular
momentum of quarks and antiquarks, and hence $2J(0)=1$ \cite{Ossmann:2004bp}.
This again is a correct and consistent result since there are no explicit gluon
degrees of freedom in the CQSM.

\begin{figure*}[t!]
\begin{tabular}{cc}
    \hspace{-1cm}
    \includegraphics[width=7cm]{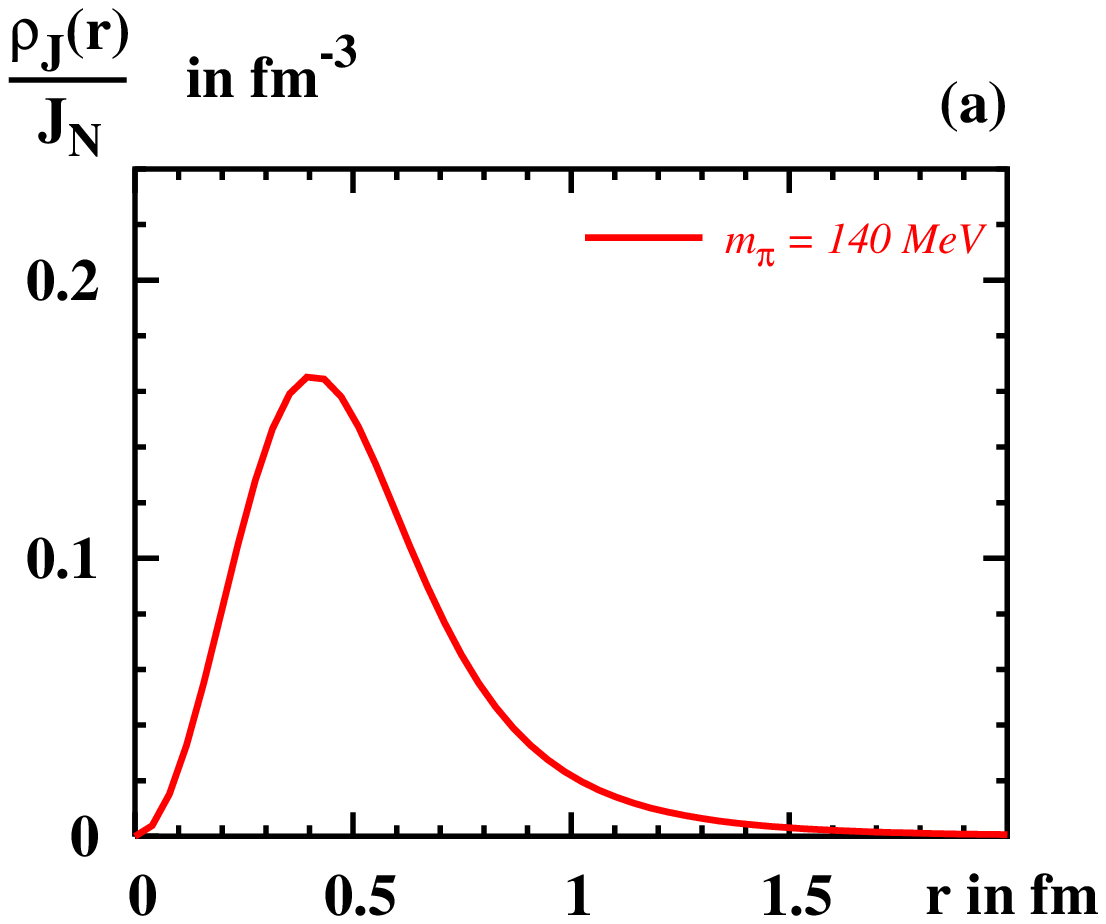}&
    \includegraphics[width=7cm]{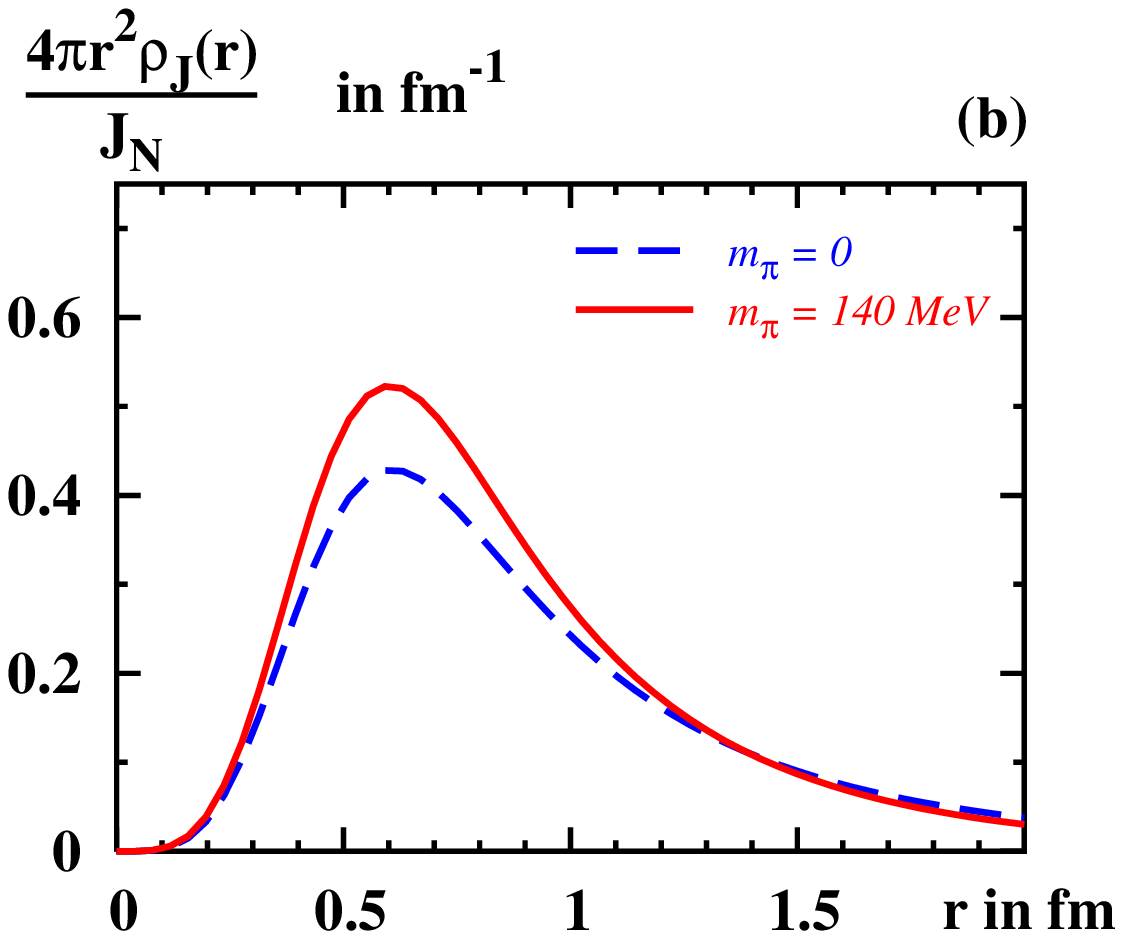}
\end{tabular}
    \caption{\label{Fig2-spin-density}
    \footnotesize\sl
    (a) The normalized angular momentum density $\rho_J(r)/J_N$ as
    function of $r$ for the physical situation with $m_\pi=140\,{\rm MeV}$.
    The curve is normalized such that it yields unity upon
    integration over the entire volume.
    (b) The normalized angular momentum density $4\pi r^2\rho_J(r)/J_N$
    as function of $r$ for $m_\pi=0$ and $140\,{\rm MeV}$. The curves are
    normalized such that one obtains unity upon integration over $r$.}
\end{figure*}

The numerical result for the normalized angular momentum density
$\rho_J(r)/J_N\equiv 2\rho_J(r)$ as function of $r$ for the physical situation
is shown in Fig.~\ref{Fig2-spin-density}a. ($J_N=\frac12$ denotes the nucleon spin.)
We observe 
that $\rho_J(r)\propto r^2$ at small $r$.

In Fig.~\ref{Fig2-spin-density}b we compare the normalized angular momentum densities
$4\pi r^2\rho_J(r)/J_N$ as functions of $r$ for $m_\pi=0$ and $140\,{\rm MeV}$.
These curves are normalized such that one obtains unity upon integration over $r$.
Within the rotating soliton picture the result is reasonable.
The smaller $m_\pi$, the larger the nucleon, and the more important is the role
of the region of large $r$ for the description of the soliton rotation, i.e.\
for the spin structure of the nucleon.
This is reflected by the mean square radius of the angular momentum density
which we define in analogy to (\ref{Eq:def-energy-mean-square-radius}) as
\be\label{Eq:def-spin-mean-square-radius}
     \la r_J^2\ra =
     \frac{\int\di^3{\bf r}\;r^2\rho_J(r)}{\int\di^3{\bf r}\;\rho_J(r)}\;.
\ee
The mean square radius of the angular momentum density increases with
decreasing $m_\pi$, see Table~\ref{Table:all-results}.
In the chiral limit $\rho_J(r) \propto \frac{1}{r^4}$ at large $r$, i.e.\
$\la r_J^2\ra$ diverges in the chiral limit. As a consequence $J(t)$ has
an infinitely steep slope at $t=0$ in the chiral limit, see
Sec.~\ref{Sec-8:form-factors} and Appendix~\ref{App:chrial-properties}
for further discussions.

\newpage

\section{\boldmath Pressure, shear forces, soliton stability, and sign of $d_1$}
\label{Sec-7:pressure+shear}

For the pressure (\ref{Eq:density-pressure}) the analogon of the ``normalization
relations'', Eqs.~(\ref{Eq:density-energy-1}) and (\ref{Eq:J-model-comp-02}),
of the other densities is the stability criterion (\ref{Eq:stability}).
Integrating the $r^2$-weighted model expression for $p(r)$ over $r$ we obtain
\be\label{Eq:stability-02}
    \int\limits_0^\infty\!\di r\,r^2 p(r)
    \equiv \frac{N_c}{12\pi}\int\di^3{\bf x}
    \sum_{n,\,\rm occ}\phi_n^\ast({\bf x})\,
    (\gamma^0{\bgam}\hat{\bf p})\,\phi_n({\bf x})=
    \frac{N_c}{12\pi} \sum_{n,\,\rm occ}\la n|\,\gamma^0
    \bgam\hat{\bf p}\,|n\ra = 0\, , \ee
because more generally the tensor $K^{ij}$ defined as
\be\label{Eq:stability-03}
    K^{ij} = \sum_{n,\,\rm occ}\la n|\,\gamma^0\gamma^i\hat{p}^j\,|n\ra
    \stackrel{\mbox{\footnotesize{Ref.~\cite{Diakonov:1996sr}}}}{=} 0\;
\ee
is zero --- however, if and only if, one evaluates the expression
(\ref{Eq:stability-03}) with the self-consistent profile, i.e.\  with
that profile which minimizes the soliton energy (\ref{Eq:soliton-energy}).
This was proven in Ref.~\cite{Diakonov:1996sr}.
Thus, the stability criterion (\ref{Eq:stability}) is satisfied.

Next let us check that one obtains in the model the correct result for the
constant $d_1=d_1(0)$ defined in Eq.~(\ref{Eq:M2-J-d1}). Expanding the
expression (\ref{Eq:d1-model-comp}) around small $t$ we obtain
\be
    d_1(t) = \frac{15 M_N}{2}\int\di^3{\bf r}\;p(r) \;
    \biggl(\frac{1}{t}+\frac{r^2}{3!} +{\cal O}(t)\biggr)\;.
        \label{Eq:d1-model-comp-02}
\ee
Due to (\ref{Eq:stability-02}) the $1/t$-term drops out,
and we verify in the CQSM the relation (\ref{Eq:d1-from-s(r)-and-p(r)})
between $d_1$ and the pressure.

\begin{figure*}[t!]
\begin{tabular}{lll}
    \hspace{-1cm}
    \includegraphics[height=5.3cm]{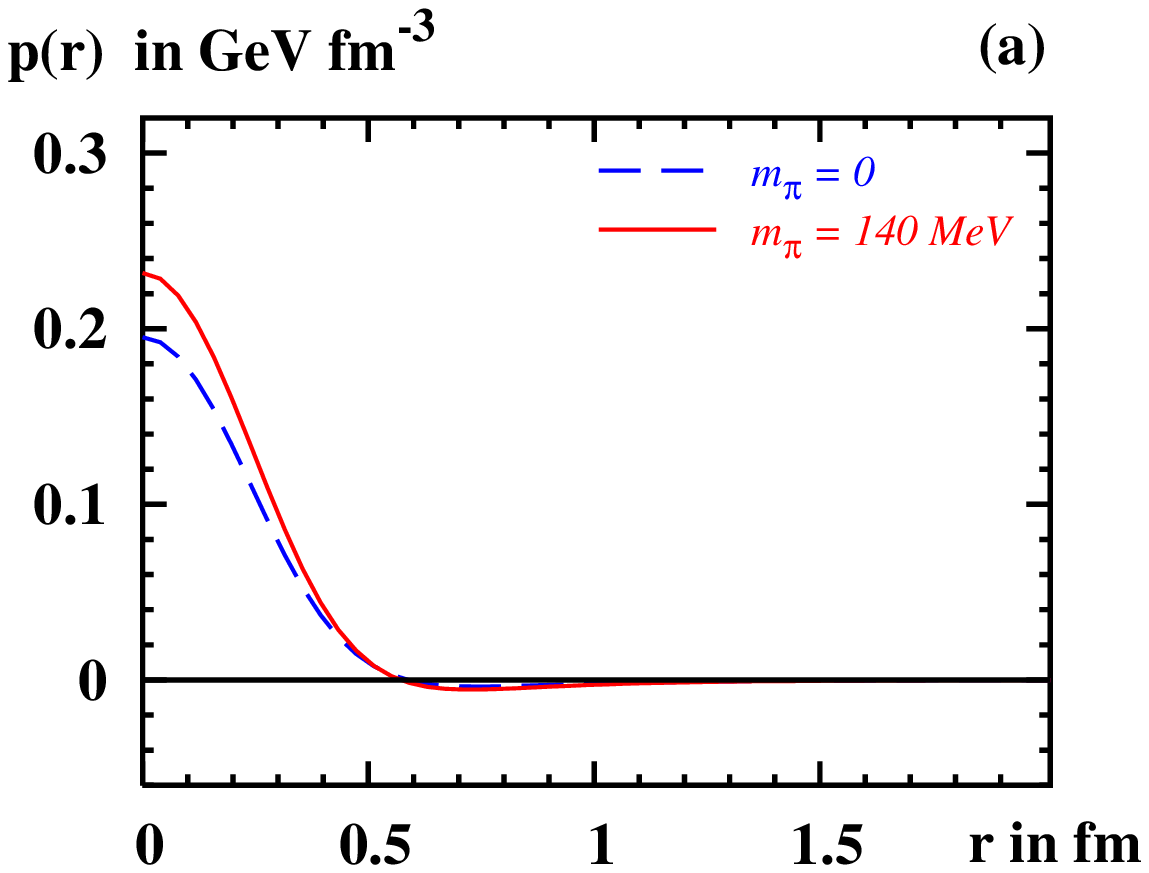}&
    \includegraphics[height=5.3cm]{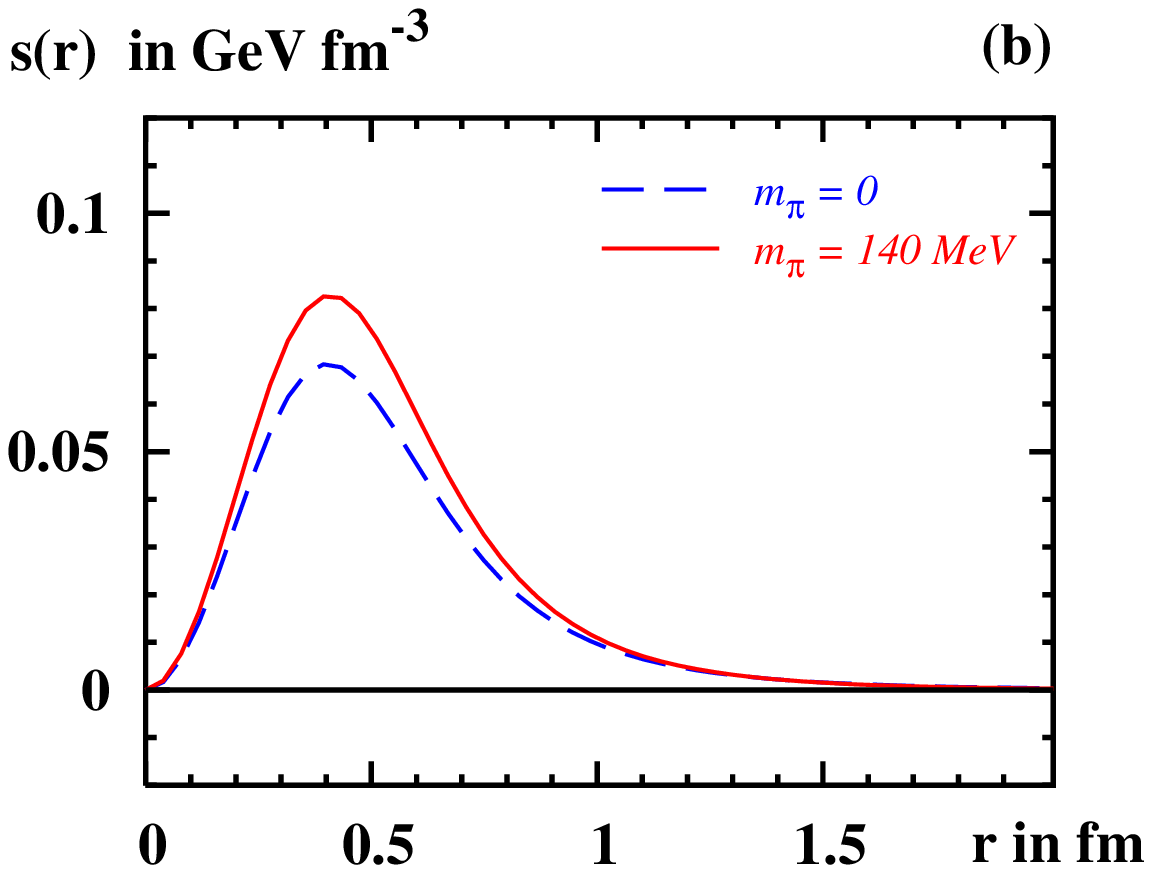} &
    \includegraphics[height=5.3cm]{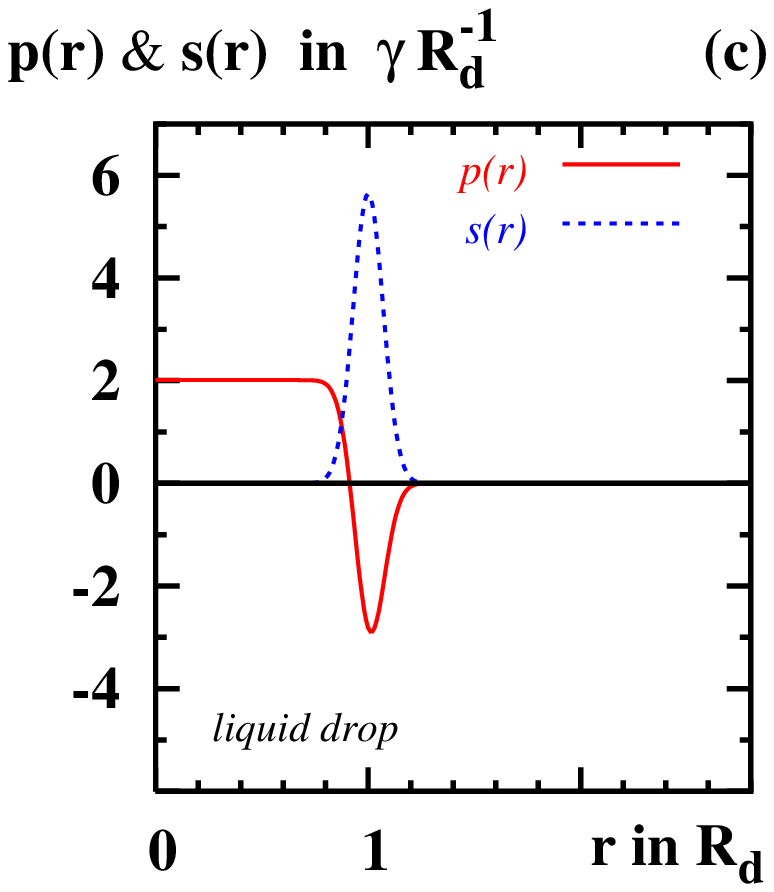}
\end{tabular}
    \caption{\label{Fig3-pressure-shear}
    \footnotesize\sl
    (a) The pressure $p(r)$ as function of $r$ for $m_\pi=0$ and
    $140\,{\rm MeV}$.
    (b) The same for the function $s(r)$ defined in
    Eq.~(\ref{Eq:T_ij-pressure-and-shear}) which describes
    the shear forces in the nucleon, and is related to $p(r)$
    by the relation (\ref{Eq:p(r)+s(r)}).
    (c) $p(r)$ and $s(r)$ in a liquid drop in units of $\gamma/R_d$
    as functions of $r$ in units of $R_d$.
    Here $R_d$ is the radius of the drop, and $\gamma$ is the surface tension.
    The $\delta$-functions in $p(r)$ and $s(r)$ in Eq.~(\ref{Eq:liquid-drop})
    are smeared for better visibility, see text.}
\end{figure*}

Fig.~\ref{Fig3-pressure-shear}a shows the pressure $p(r)$ as function of $r$.
In the physical situation $p(r)$ takes its global maximum at $r=0$ with
$p(0) = 0.23\,{\rm GeV}/{\rm fm}^3 = 3.7\cdot10^{34}\,{\rm Pa}$.
This is ${\cal O}(10\!-\!100)$ higher than the pressure inside a neutron star
\cite{Prakash:2000jr}.
Then $p(r)$ decreases monotonically --- becoming zero at $r_0 = 0.57\,{\rm fm}$ ---
till reaching its global minimum at $r_{p,\,\rm min}= 0.72\,{\rm fm}$, after which
it increases monotonically remaining, however, always negative.
The positive sign of the pressure for $r<r_0$ corresponds to repulsion,
while the negative sign in the region $r > r_0$ means attraction.

In Fig.~\ref{Fig3-pressure-shear}a we see how the pressure depends on the pion mass.
The pressure in the center of the nucleon increases as $m_\pi$ increases --- obvious
consequence of the fact that the (energy) density also increases,
see Table~\ref{Table:all-results}.
As a response to the increased $p(r)$ at small $r$ --- keep in mind the
stability condition (\ref{Eq:stability}) ---  the pressure takes
also larger absolute values in the region $r>r_0$ where it is negative.
This can again be intuitively understood because a heavier particle is more
tightly bound, i.e.\ the attractive forces are stronger.
The zero of $p(r)$ moves towards smaller values of $r$ with increasing $m_\pi$,
see Table~\ref{Table:all-results}.

In Fig.~\ref{Fig3-pressure-shear}b we show the distribution
of the shear forces $s(r)$ obtained from our results for $p(r)$ by
solving the differential equation (\ref{Eq:p(r)+s(r)}).
The  distribution of shear forces is always positive. It reaches
for $m_\pi=140\,{\rm MeV}$ a global maximum at $r=0.40\,{\rm fm}$.
The position of the maximum is weakly dependent on $m_\pi$.
At small $r$ we observe $s(r)\propto r^2$.

It is interesting to compare to which extent the nucleon ``resembles'' a liquid
drop of radius $R_d$ with constant density and constant pressure $p_0$ inside.
In such a drop the pressure and shear forces are given by \cite{Polyakov:2002yz}
\be\label{Eq:liquid-drop}
    p(r) = p_0\,\theta(R_d-r)-\frac13\;p_0R_d\,\delta(R_d-r)
    \;\;\;\;\mbox{and}\;\;\;\;
    s(r) = 
           \gamma\,\delta(R_d-r)\;,\ee
where $\gamma=\frac12\;p_0R_d$ denotes the surface tension.
We show this situation in Fig.~\ref{Fig3-pressure-shear}c --- where,
however, for better visibility the $\delta$-functions in (\ref{Eq:liquid-drop})
are smeared out. This corresponds to allowing the density in the drop
to decrease continuously from its constant inner value to zero over
a finite ``skin'' (of the size $\sim\frac{1}{10}R_d$ in
Fig.~\ref{Fig3-pressure-shear}c).

Comparing the liquid drop picture to the results from the CQSM we observe
a remote qualitative similarity.
In contrast to the liquid drop, the density ``inside'' the nucleon is far from
being constant, see Fig.~\ref{Fig1-energy-density}a, and one cannot expect the
pressure in the nucleon to exhibit a constant plateau as in the liquid drop.
Still the pressure exhibits the same qualitative features.
The shear forces become maximal in the vicinity of what can be considered as
the ``edge'' of the object. This is the case in particular for the liquid drop
However, the ``edge'' of nucleon is far more diffuse, and the distribution of
shear forces $s(r)$ is widespread.
Of course, the nucleon can hardly be considered a liquid drop.
Such an analogy might be more appropriate for nuclei \cite{Polyakov:2002yz}.
Nevertheless this comparison gives some intuition on the model results ---
in particular, about the qualitative shape of the distributions of pressure
and shear forces.

\begin{figure*}[t!]
\begin{tabular}{ll}
  \hspace{-0.5cm}
  \includegraphics[height=5.3cm]{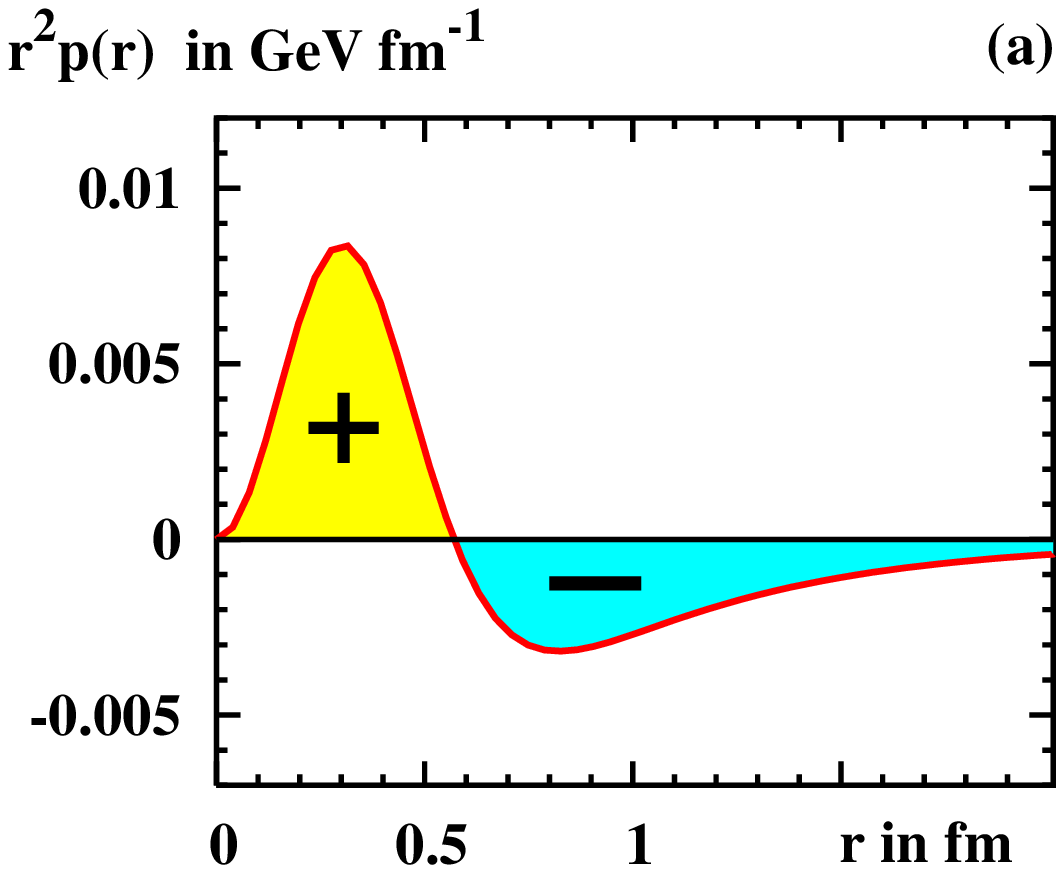}&
  \includegraphics[height=5.3cm]{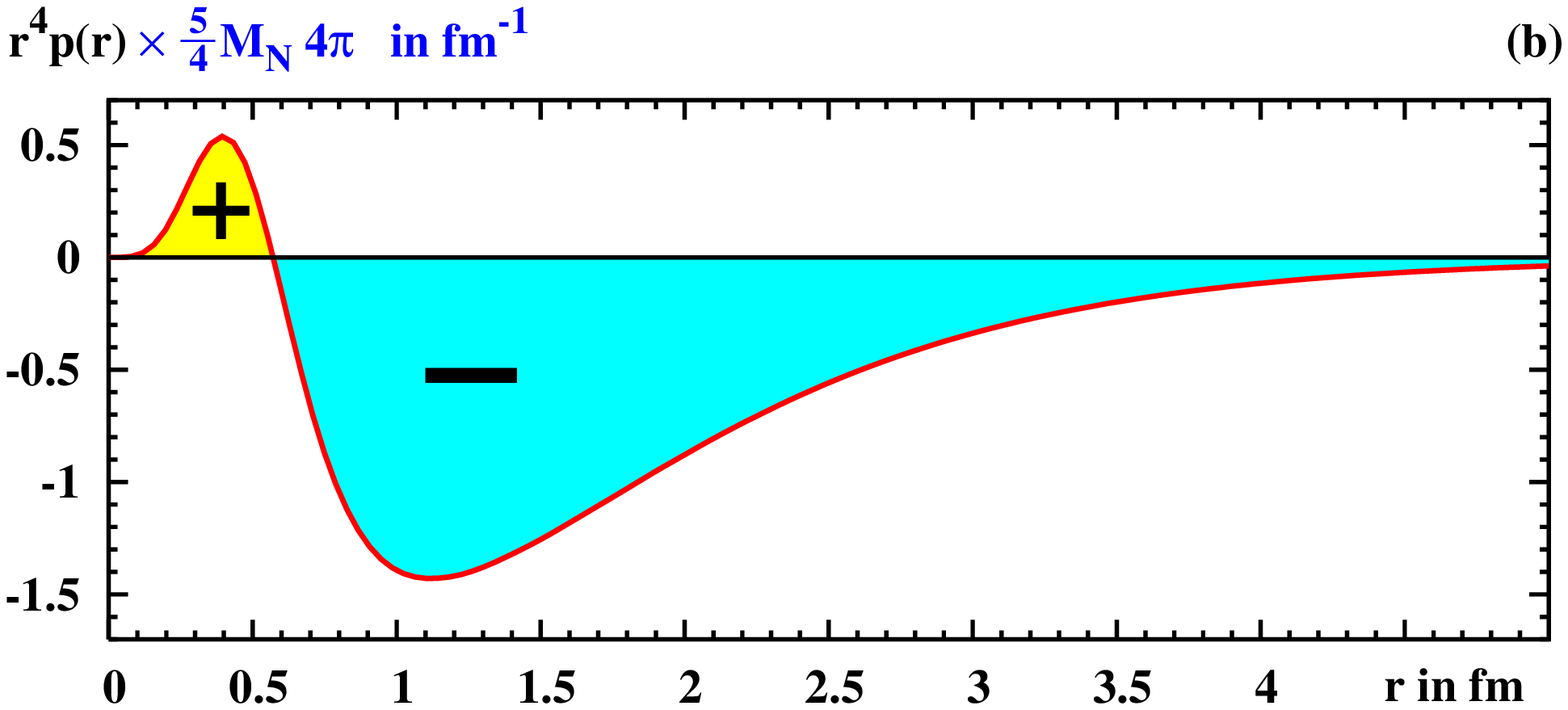}
\end{tabular}
    \caption{\label{Fig7-pressure-sign-d1}
    \footnotesize\sl
    (a)
    $r^2 p(r)$ as function of $r$ from the CQSM at the physical value of $m_\pi$.
    The shaded regions have --- within the numerical accuracy of about
    half percent --- the same surface areas. This shows how the stability
    condition $\int_0^\infty\di r\,r^2p(r)=0$ in Eq.~(\ref{Eq:stability-02})
    is realized.\\
    (b)
    The same as (a) but with an additional power of $r^2$ and the prefactor
    $5\pi M_N$. Integrating this curve over $r$ yields $d_1$ according to
    (\ref{Eq:d1-from-s(r)-and-p(r)}).
    The plot shows that one obtains a negative sign for $d_1$
    as a consequence of the stability condition (\ref{Eq:stability})
    shown in Fig.~\ref{Fig7-pressure-sign-d1}a.}
\end{figure*}

Next let us discuss how the stability condition (\ref{Eq:stability-02}) is satisfied.
Fig.~\ref{Fig7-pressure-sign-d1}a shows $r^2p(r)$ as function of $r$.
The shaded regions have the same surface areas but opposite sign
and cancel each other --- within numerical accuracy
\be\label{Eq:pressure-02}
    \int\limits_0^{r_0} \di r\;r^2p(r) =  2.61\,{\rm MeV}\,,\;\;\;\;
    \int\limits_{r_0}^\infty\di r\;r^2p(r) = -2.63\,{\rm MeV}\,.
\ee

\begin{wrapfigure}[18]{R}{8cm}
\vspace{-1cm}
\includegraphics[width=8cm]{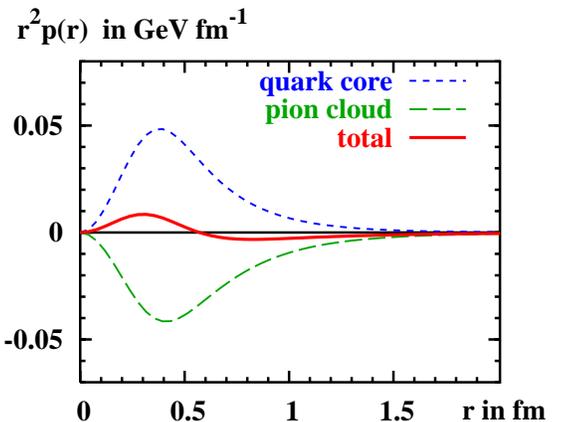}
    \caption{\label{Fig5-pressure-details}
    \footnotesize\sl
    The pressure $p(r)$ as function of $r$ for $m_\pi=140\,{\rm MeV}$.
    Dotted line: Contribution of the discrete level associated with
    the quark core. Dashed line: Continuum contribution associated with
    the pion cloud. Solid line: The total result.}
\end{wrapfigure}

In order to better understand how the soliton acquires stability,
it is instructive to look in detail how the total pressure is decomposed of the
separate contributions of the discrete level and the continuum contribution.
Fig.~\ref{Fig5-pressure-details} shows that the contribution of the
discrete level is always positive.
This contribution corresponds in model language to the contribution
of the ``quark core'' and one expects a positive contribution
(``repulsion'') due to the Pauli principle.
At large $r$ the discrete level contribution vanishes exponentially
since the discrete level wave-function does so \cite{Diakonov:1987ty}.

The continuum contribution is throughout negative --- as can be seen from
Fig.~\ref{Fig5-pressure-details} and can be understood as follows.
The continuum contribution can be interpreted as the effect of the pion cloud
which in the model is responsible for the forces binding the quarks to form the nucleon.
I.e.\  it provides a negative contribution to the pressure corresponding to attraction.
In the chiral limit the continuum contribution exhibits a power-like decay which
dictates the long-distance behaviour of the total result for the pressure as
follows
\be\label{Eq:pressure-03}
    p(r) = - \left(\frac{3g_A}{8\pi f_\pi}\right)^2 \frac{1}{r^6}
    \;\;\;\;\mbox{and}\;\;\;\;
    s(r) = 3 \left(\frac{3g_A}{8\pi f_\pi}\right)^2 \frac{1}{r^6}\;\;\;\;\;\;
    \mbox{at large $r$.}
\ee
where for completeness we quote also the result for $s(r)$.
For $m_\pi\neq 0$ the continuum contribution exhibits an exponential decay at
large $r$ due to the Yukawa tail of the soliton profile (\ref{Eq:profile-at-large-r})
--- like the contribution of the discrete level, however, it still dominates the
behaviour of the total result.
Fig.~\ref{Fig5-pressure-details} reveals that the actual cancellation between
the different contributions leading to (\ref{Eq:stability-02}) are even more
impressive than we could guess considering the numbers in Eq.~(\ref{Eq:pressure-02}).
For $m_\pi=140\,{\rm MeV}$ we obtain the following numbers, and see
that the two contributions cancel each other within numerical accuracy
\be\label{Eq:pressure-04}
    \int\limits_0^\infty\di r\;r^2p(r)\biggl|_{\rm lev} = 26.69\,{\rm MeV}\,,
    \;\;\;\;
    \int\limits_0^\infty\di r\;r^2p(r)\biggl|_{\rm cont}=-26.71\,{\rm MeV}\,.
\ee

Finally, we discuss the relation of stability and the sign of the constant $d_1$.
By comparing Figs.~\ref{Fig7-pressure-sign-d1}a and \ref{Fig7-pressure-sign-d1}b
we immediately understand that in the CQSM the constant $d_1$ takes a negative value
\be\label{Eq:d1-sign}
    d_1 < 0\,.
\ee
This result is presumably of general character.
In fact, the observation (\ref{Eq:d1-sign}) follows naturally from our intuition on
the pressure and shear forces distributions we gained from our study.
It seems physically intuitive that in a mechanically stable object the
following conditions hold
\ba
    \mbox{(I)}
    &&
    \mbox{for a stable object the pressure has the shape}\;\;\;\;
    \displaystyle p(r)\;\cases{> 0 & for $r<r_0$\cr
                       = 0 & for $r=r_0$\cr
                   < 0 & for $r>r_0$}\nonumber\\
    \phantom{X}\nonumber\\
    \mbox{(II)}
&&  \mbox{distribution of shear forces satisfies for all $r$}\;\;\;\;\;\;
    s(r)>0\;.
    \nonumber\ea
From (I) and the stability criterion (\ref{Eq:stability})
we conclude that $d_1\propto \int_0^\infty \di r \,r^4p(r)<0$ as illustrated
in Figs.~\ref{Fig7-pressure-sign-d1}a and \ref{Fig7-pressure-sign-d1}b.
The same conclusion follows from (II) using
$d_1\propto -\int_0^\infty \di r \,r^4s(r)<0$ in (\ref{Eq:d1-from-s(r)-and-p(r)}).

Of course, one can imagine a pressure distribution with more zeros
and a different shape than in (I) still yielding
(\ref{Eq:d1-sign}). However, (I) is the simplest case which one
may expect to hold for a ground state --- like the nucleon. Such a
ground state object is characterized by having one ``surface''
only, although a quite smeared out one in the case of the nucleon,
hence the condition (II). A more complicated distribution of the
pressure --- with more zeros, i.e.\ also more maxima and minima
--- would imply an object with several (smeared out) surfaces,
which follows from the last condition in
App.~\ref{App:general-relations-from-EMT-conservation}.

The conjecture (\ref{Eq:d1-sign}) is --- besides being physically appealing---
in agreement with all available information on $d_1$, see
Sec.~\ref{Sec-2:FF-of-EMT-in-general}. One may therefore suspect that
(\ref{Eq:d1-sign}) is a general theorem which connects the stability of an
object to the sign of its constant $d_1$.
However, such a theorem --- if it exists --- remains to be rigorously
proven for the general case.

\newpage
\section{\boldmath Results for the form factors}
\label{Sec-8:form-factors}

From the densities $\rho_E(r)$, $\rho_J(r)$ and $p(r)$ which we discussed in
detail in Secs.~\ref{Sec-5:energy-density}, \ref{Sec-6:spin-density} and
\ref{Sec-7:pressure+shear} we obtain the form factors of the EMT by means
of Eqs.~(\ref{Eq:M2-d1-model-comp},~\ref{Eq:d1-model-comp},~\ref{Eq:J-model-comp}).
The results are shown in Fig.~\ref{Fig7-ffs} for $m_\pi=0$ and $140\,{\rm MeV}$.
In the CQSM the form factors $M_2(t)$, $J(t)$ are normalized at
$t=0$ as $M_2(0)=2J(0)=1$ as proven in
Secs.~\ref{Sec-5:energy-density} and \ref{Sec-6:spin-density}.
The numerical results satisfy these constraints within a numerical
accuracy of (1-2)$\%$, see Figs.~\ref{Fig7-ffs}a and \ref{Fig7-ffs}b.

In contrast, the normalization of the form factor $d_1(t)$ at $t=0$ is not known
a priori. We find $d_1=d_1(0)<0$ as anticipated in Sec.~\ref{Sec-7:pressure+shear}.
The constant $d_1$ has a well-defined chiral limit. This can be concluded from
Eq.~(\ref{Eq:d1-from-s(r)-and-p(r)}) which relates $d_1$ to the distributions of
pressure or shear forces, and from the fact that $p(r)$ and $s(r)$ drop off
sufficiently fast at large $r$ in the chiral limit, see (\ref{Eq:pressure-03}).

\begin{wrapfigure}[18]{R}{8cm}
\vspace{-1cm}
\includegraphics[width=8cm]{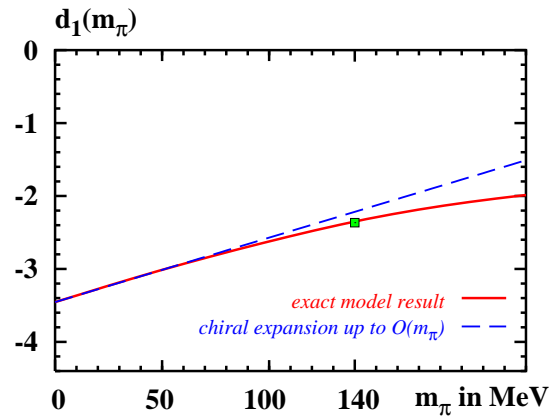}
    \caption{
    \footnotesize\sl
    The constant $d_1$ as function of $m_\pi$.
    Comparison of the full CQSM result, and the expansion
    of $d_1(m_\pi)$ up the leading non-analytic term according to
    Eq.~(\ref{Eq:d1-vs-mpi}). The square marks the physical point.}
\end{wrapfigure}

The absolute value of $d_1$ decreases with increasing $m_\pi$ and we observe a strong
sensitivity of $d_1$ to $m_\pi$. That this is not surprizing can be understood by
considering the chiral expansion of $d_1$. Expanding $d_1(m_\pi)$ in the model for
small $m_\pi$ we obtain, cf. Appendix~\ref{App:chrial-properties},
\be\label{Eq:d1-vs-mpi}
    d_1(m_\pi)
    = \;\krig{d_1}\;+\; \frac{15\,g_A^2M_N}{64\,\pi\,f_\pi^2}\;m_\pi
    + \dots
\ee
Thus, we see that $d_1(m_\pi)$ receives a large leading
non-analytic contribution in the current quark mass $m \propto m_\pi^2$,
which is the origin of the strong $m_\pi$-dependence of $d_1(m_\pi)$.
The expansion (\ref{Eq:d1-vs-mpi}) approximates in the model $d_1(m_\pi)$
to within an accuracy of $15\%$ up to the physical point.
At the physical point the absolute value of $d_1$ is reduced by about $30\,\%$
with respect to its chiral limit value. The numerical results for
$d_1$ are summarized in Table~\ref{Table:all-results}.

Next, let us focus on the derivatives of the form factors at $t=0$.
From Eq.~(\ref{Eq:ff-M2}) we conclude that at $t=0$ the slope
of the form factor $M_2(t)$ is given by
\be\label{Eq:M2-slope-at-t=0}
    M_2^\prime(0) = \frac{\la r_E^2\ra}{6} + \frac{d_1}{5M_N^2}
    = \krig{M}{\!}_2^\prime(0)
    - \frac{21\,g_A^2}{128  \pi f_\pi^2}\; \frac{m_\pi}{M_N}
    + \dots \ee
where the chiral expansion follows from (\ref{Eq:rE2-vs-mpi}) and
(\ref{Eq:d1-vs-mpi}). In particular, $M_2^\prime(0)$ is well-defined
for all $m_\pi$ including the chiral limit.
The situation is different for the form factors $J(t)$ and $d_1(t)$.
The slope of $J(t)$ at $t=0$ can be deduced from (\ref{Eq:J-model-comp})
and is given by
\be\label{Eq:J-slope-at-t=0}
    J^\prime(0) = \frac{\la r_J^2\ra}{6}
\ee
with $\la r_J^2\ra$ defined in (\ref{Eq:def-spin-mean-square-radius}). Since the
mean square radius of the angular momentum density diverges in the chiral limit,
see Sec.~\ref{Sec-6:spin-density}, so does $J^\prime(0)$.
See also the discussion in App.~\ref{App:chrial-properties}.

That also the slope of $d_1(t)$ becomes infinitely steep at $t=0$
can be understood as follows.
From Eqs.~(\ref{Eq:ff-M2},~\ref{Eq:T_ij-pressure-and-shear},~\ref{Eq:p(r)+s(r)})
it follows that the derivative of $d_1(t)$ at $t=0$ be expressed as
\be\label{Eq:d1-slope-at-t=0}
    d_1^{\,\prime}(t)\biggl|_{t=0}
    = -\frac{M_N}{42}\,\int\di^3{\bf r}\;r^4s(r)
    =  \frac{M_N}{16}\,\int\di^3{\bf r}\;r^4p(r)\;.
\ee
Since in the chiral limit $p(r)$ and $s(r)$ drop off as $\frac{1}{r^6}$
at large $r$ we see that $\krig{d_1^{\,\prime}}\!(0)$ diverges as $m_\pi\to 0$.

To make this statement more quantitative let us expand $d_1(t)$
in the chiral limit in the model for small $t$. We obtain
\be\label{Eq:d1-at-small-t-chir-lim}
    \krig{d_1}\!(t) = \krig{d_1} -
    \frac{45\,g_A^2M_N}{512\,f_\pi^2}\;\sqrt{-t}+{\cal O}(t)\;,
\ee
which means that in the chiral limit $\krig{d_1^{\,\prime}}\!(t)\propto 1/\sqrt{-t}$
at small $t$.
Alternatively, one may keep $m_\pi\neq 0$, evaluate the derivative of $d_1(t)$ at
$t=0$, and then consider small $m_\pi$.
(Note that the limits $m_\pi\to0$ and $t\to 0$ do not commute.)
Then we find that the slope of $d_1(t)$ at $t=0$ diverges in the chiral limit as
\be\label{Eq:d1-slope-at-t=0-chir-lim}
    \krig{d_1^{\,\prime}}\!(t)\biggl|_{t=0} =
    -\frac{3\,g_A^2 \,M_N}{32\, \pi f_\pi^2 \,m_\pi} + \dots \;.
\ee
The results (\ref{Eq:d1-slope-at-t=0-chir-lim}) and
(\ref{Eq:d1-slope-at-t=0-chir-lim}) are derived in
Appendix~\ref{App:chrial-properties}.
The numerical results for $J(t)$ and $d_1(t)$ in Figs.~\ref{Fig7-ffs}b and
\ref{Fig7-ffs}c indicate the infinitely steep slopes at $t=0$ within the
numerical accuracy.

Notice that we derived the analytical results in (\ref{Eq:d1-slope-at-t=0},
\ref{Eq:d1-at-small-t-chir-lim}, \ref{Eq:d1-slope-at-t=0-chir-lim}) in the
framework of the CQSM. However, the leading non-analytic terms
(i.e.\ terms $\propto m_\pi$) in the chiral expansion of $d_1(t)$
in (\ref{Eq:d1-slope-at-t=0}, \ref{Eq:d1-at-small-t-chir-lim},
\ref{Eq:d1-slope-at-t=0-chir-lim})
are dictated by chiral symmetry breaking, and are {\sl independent} of the
details of the chiral theory chosen to derive them.
In fact, our results (\ref{Eq:d1-slope-at-t=0}, \ref{Eq:d1-at-small-t-chir-lim},
\ref{Eq:d1-slope-at-t=0-chir-lim}) agree with those obtained from chiral
perturbation theory in Ref.~\cite{Belitsky:2002jp}
--- provided one takes into account an important difference.
The CQSM is formulated in the large-$N_c$ limit which does not commute
with the chiral limit \cite{Dashen:1993jt}. At large $N_c$ the masses of the nucleon
and $\Delta$-resonance are degenerated: $M_\Delta-M_N={\cal O}(N_c^{-1})$.
Therefore in the CQSM --- in addition to the nucleon considered in chiral
perturbation theory \cite{Belitsky:2002jp} --- the $\Delta$-resonance contributes on
equal footing as intermediate state in chiral loops. Considering that in the large
$N_c$-limit the pion-Delta-nucleon and pion-nucleon-nucleon couplings are related
as $g_{\pi N\Delta}=\frac32 g_{\pi NN}$
(phenomenologically satisfied to a very good approximation),
one finds that the $\Delta$-resonance makes a contribution to leading non-analytic
terms which is --- for scalar-isoscalar quantities ---two times larger than that
of the nucleon \cite{Cohen:1992uy}. Hence, our leading non-analytic terms in
Eqs.~(\ref{Eq:d1-slope-at-t=0}, \ref{Eq:d1-at-small-t-chir-lim},
\ref{Eq:d1-slope-at-t=0-chir-lim}) are 3 times larger than those obtained from
chiral perturbation theory in Ref.~\cite{Belitsky:2002jp} where $N_c$ was kept
finite.\footnote{
    Actually, there is one more subtlety to be considered. In the result for the
    form factor $d_1(t)=\frac45C_2(t)$ in Eq.~(20) of \cite{Belitsky:2002jp},
    see App.~\ref{App:Alternative-definition} for the discussion of the notation,
    in addition the constant $a_2^{Q,\pi}\equiv M_2^{Q,\pi}$ appears,
    which describes the fraction of the pion momentum carried by quarks.
    In the effective theory (\ref{eff-theory}) quarks carry the entire
    momentum of the pion, i.e.\ $M_2^{Q,\pi}=1$ \cite{Polyakov:1999gs}.}
Other examples of the derivation of leading non-analytic terms in
chiral soliton models can be found in \cite{Schweitzer:2003sb,Cohen:1992uy}.

\begin{figure*}[t!]
\begin{tabular}{ccc}
  \hspace{-0.5cm}
  \includegraphics[width=6cm]{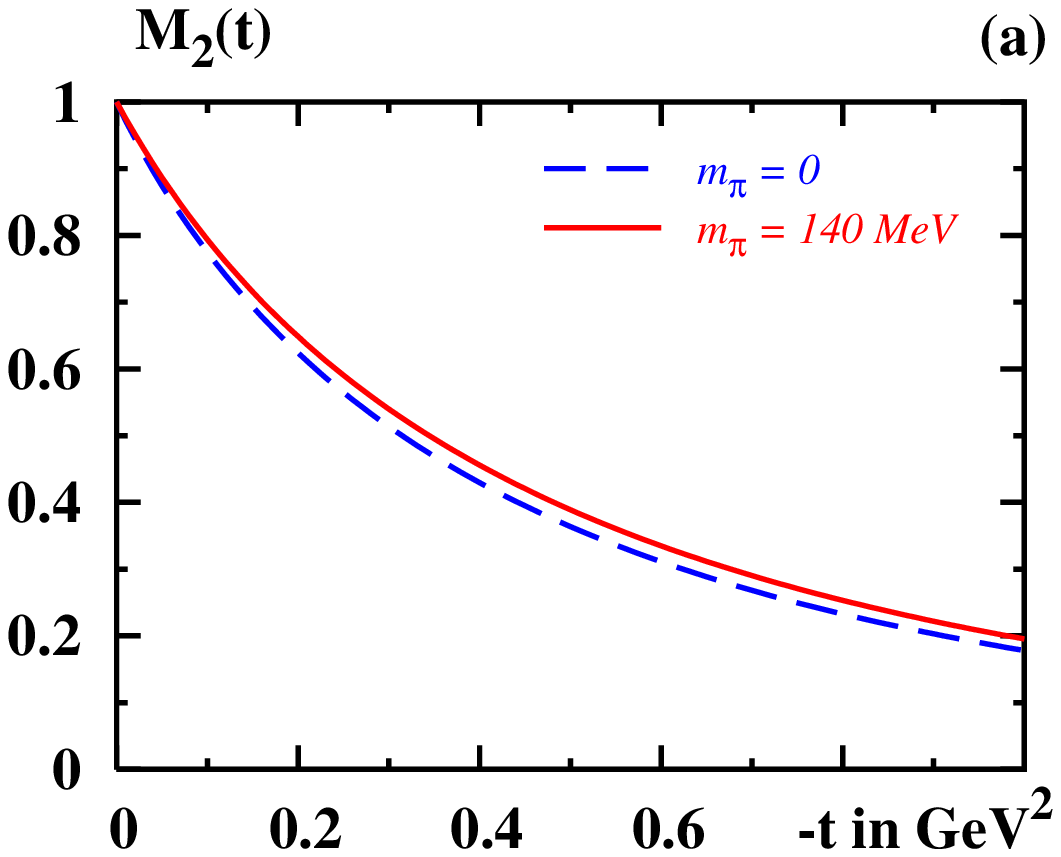}&
  \includegraphics[width=6cm]{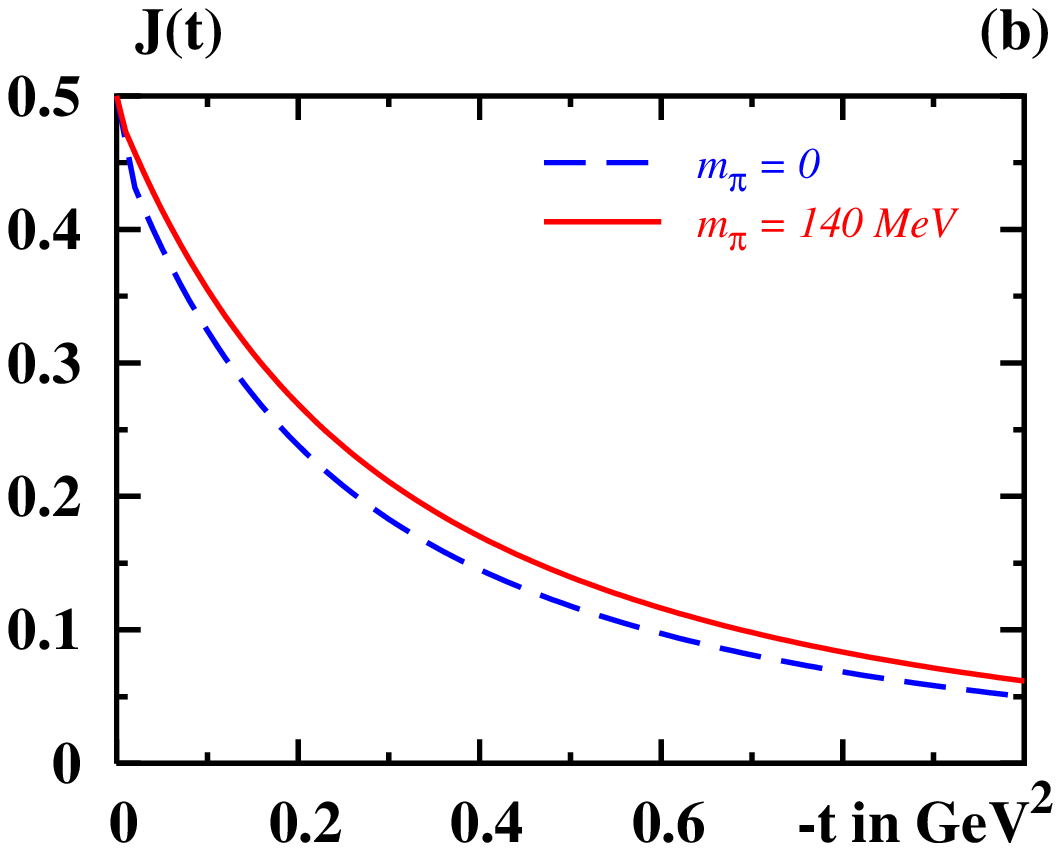}&
  \includegraphics[width=6cm]{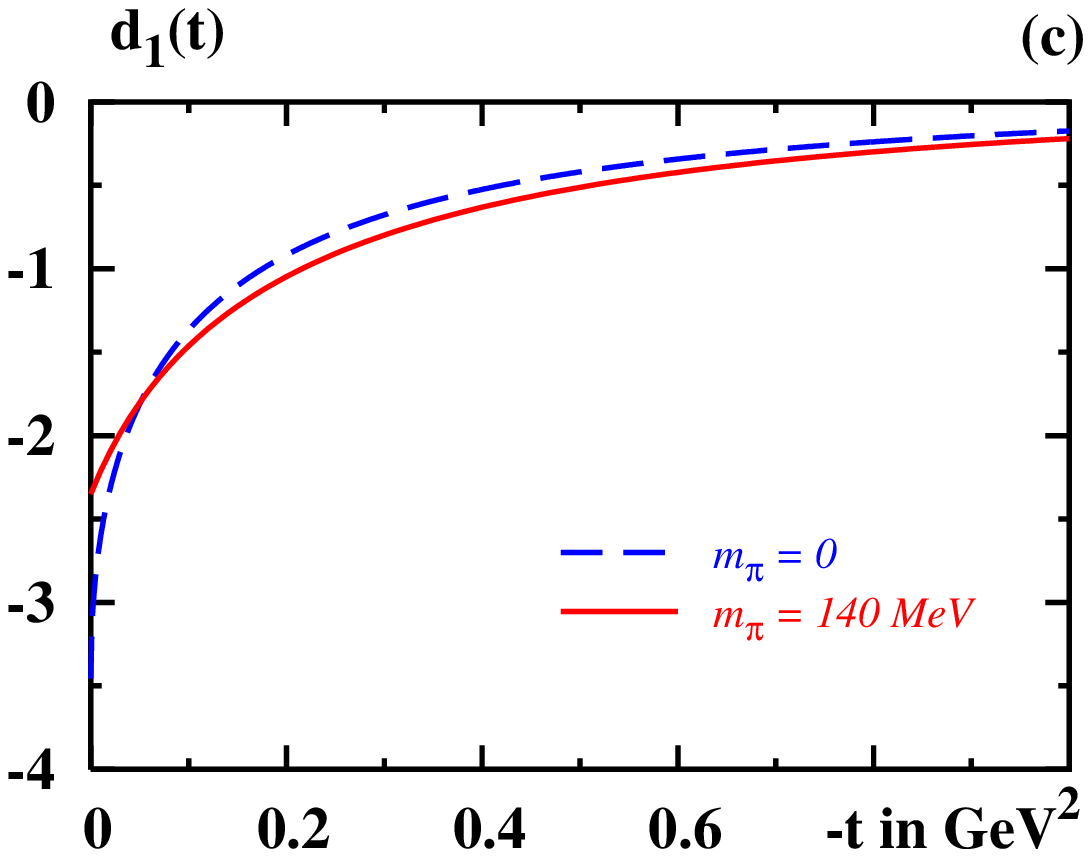}
\end{tabular}
    \caption{\label{Fig7-ffs}
    \footnotesize\sl
    The form factors of the energy momentum tensor
    $M_2(t)$, $J(t)$ and $d_1(t)$ as functions of $t$ for
    the pion masses $m_\pi = 0$, $140\,{\rm MeV}$.
    All form factors can be well approximated by dipole fits, however,
    with the exception of $J(t)$ and $d_1(t)$ in the chiral limit
    which exhibit infinitely steep slopes at $t=0$, see text.}
\end{figure*}

\begin{wrapfigure}[18]{R}{8.3cm}
\vspace{-1cm}
  \begin{center}
  \includegraphics[width=7.3cm]{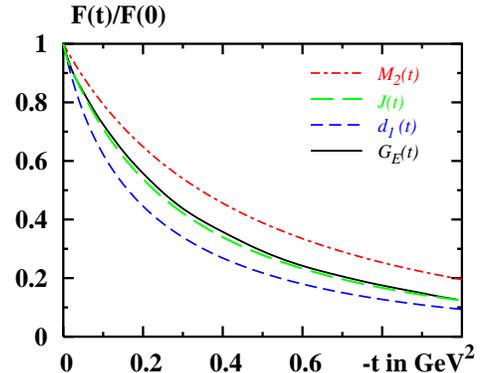}
  \end{center}
\vspace{-0.5cm}
        \caption{\label{Fig8-comparing-FFs}
        \footnotesize\sl
    EMT form factors as functions of $t$.
    Dashed-dotted line: $M_2(t)$.
    Long-dashed line: $J(t)$.
    Dashed line: $d_1(t)$; all computed here.
    Solid line: electric proton form factor $G_E(t)$
    from Ref.~\cite{Christov:1995hr}.
    All form factors refer to the physical pion mass, and are normalized
    with respect to their values at $t=0$.}
\vspace{-0.5cm}
\end{wrapfigure}

Next, we turn to the discussion of the $t$-dependence of the form factors, and
recall that in the large-$N_c$ limit we strictly speaking are restricted to
$|t|\ll M_N^2$.
However,  in practice it is observed that the CQSM provides reliable results
for electromagnetic form factors up to $|t|\lesssim 1\,{\rm GeV}^2$
\cite{Christov:1995hr,Christov:1995vm}.

Fig.~\ref{Fig7-ffs} shows the form factors of the EMT as functions of $t$ for
$|t|\le 1\,{\rm GeV}^2$ for $m_\pi=0$ and $140\,{\rm MeV}$.
For $m_\pi \neq 0$ the form factors can be well approximated by dipole fits
\be\label{Eq:dipol}
    F(t) = \frac{F(0)}{(1-t/M_{\rm dip}^2)^2}
\ee
with the values for the dipole masses quoted in Table~\ref{Table:all-results}.

It is instructive to compare these results within the model to the electromagnetic
form factors --- for definiteness we choose the electric form factor of the proton
$G_E(t)$ computed in the CQSM in Ref.~\cite{Christov:1995hr}.
Fig.~\ref{Fig8-comparing-FFs} shows that $J(t)$ and $G_E(t)$ exhibit a similar
$t$-dependence. However, $M_2(t)$ falls off with increasing $|t|$ slower than
$G_E(t)$, while $d_1(t)$ shows a faster fall off.

One popular assumption in literature in the context of modelling GPDs is
to assume a generic factorized Ansatz of the type $H(x,\xi,t)=H(x,\xi)G(t)$
where $G(t)$ denotes the respective form factor
(other approaches are discussed in \cite{Goeke:2001tz,Belitsky:2001ns,Diehl:2004cx,Guidal:2004nd,Radyushkin:2004mt,Vanderhaeghen:2004sa}).
This assumption implies that the form factors of the EMT should
have approximately the same $t$-dependence as the electromagnetic form factors.
Our results indicate that this is quite a rough approximation,
and support the observations that in the CQSM
$H(x,\xi,t)\neq H(x,\xi)G(t)$, see Refs.~\cite{Goeke:2001tz,Petrov:1998kf}

Let us compare our result for $d_1$ with the result from direct
calculations of GPDs in the model \cite{Petrov:1998kf} which
yield\footnote{
    From this number it was estimated that $d_1\approx -4.0$
    at experimentally relevant scales of few ${\rm GeV}^2$
    \cite{Kivel:2000fg} in the following way. The model predicts
    for the {\sl ratio} $d_1^Q/M_2^Q \approx -8.0$ and experimentally
    one finds $M_2^Q\approx 0.5$ at scales of few ${\rm GeV}^2$.
    This estimate neglects strictly speaking the different evolution
    properties of $d_1^Q$ and $M^Q$ which is justified because
    uncertainties in the model dominate.}
$d_1^Q \approx -8.0$ at the low model scale in the chiral limit.
The discrepancy with the corresponding results in Table~\ref{Table:all-results}
is due to the fact that in \cite{Petrov:1998kf} the momentum-dependent constituent
mass $M(p)$ was employed vs.\ our proper-time or Pauli-Villars regularization with
$M={\rm const}$, and that the approximative ``interpolation formula'' was used
vs.\ the exact numerical calculation done here.

\begin{wrapfigure}[13]{R}{7.8cm}
\vspace{-1.5cm}
\begin{center}
\includegraphics[width=7.3cm]{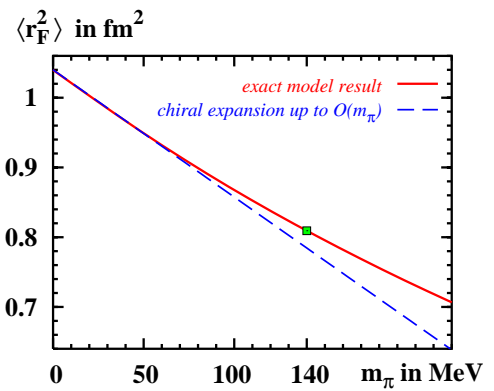}
\end{center}
\vspace{-0.8cm}
        \caption{\label{Fig9-rF2-vs-mpi}
        \footnotesize\sl
    Mean square radius of EMT trace operator
    vs.\ $m_\pi$.}
\end{wrapfigure}

For the mean square radius $\la r_F^2\ra$ of the operator of the EMT trace,
see Eqs.~(\ref{Eq:ff-of-EMT-trace-4},~\ref{Eq:ff-of-EMT-trace-5}),
we obtain in the chiral limit
\be\label{Eq:ff-of-EMT-trace-10}
    \la\krig{r}{\!}_F^2\ra = (1.0-1.2)\,{\rm fm}^2
\ee
while its value at the physical point is reduced by about $0.2\,{\rm fm}$
compared to (\ref{Eq:ff-of-EMT-trace-10}), see Table~\ref{Table:all-results}.
From the chiral expansion of $\la r_F^2\ra$ which follows from
Eqs.~(\ref{Eq:rE2-vs-mpi},~\ref{Eq:d1-vs-mpi},~\ref{Eq:ff-of-EMT-trace-5}) and reads
\be\label{Eq:rF2-chiral-expansion}
    \la r_F^2\ra =
    \la\krig{r}{\!}_F^2\ra - \frac{117\,g_A^2}{64 \pi f_\pi^2 M_N}\; m_\pi + \dots
    \;,
    \ee
we see that the leading non-analytic term in (\ref{Eq:rF2-chiral-expansion})
explains the main portion of the observed reduction of $\la r_F^2\ra$ at the physical
point compared to its chiral limit value, see Fig.~\ref{Fig9-rF2-vs-mpi}.

It is instructive to compare our
result (\ref{Eq:ff-of-EMT-trace-10}) for the mean square radius of
the EMT trace to the mean square radius of the traceless part of
the EMT $(0.3\,{\rm fm})^2$ estimated by means of QCD sum rules
\cite{Braun:1992jp}. The instanton vacuum model provides a
possible explanation why the radius of the trace of the EMT is so
much larger than the traceless part.
In chiral limit the trace part is basically the gluonic
operator $F_{\mu\nu}F^{\mu\nu}$ which is due to one-instanton
contributions and appears in leading order in the
instanton packing fraction. The traceless part arises from
instanton-anti-instanton contributions which appear at subleading
order \cite{Diakonov:1995qy}.

%
\begingroup
\squeezetable
\begin{table}[t!]
    \caption{\footnotesize\sl
    \label{Table:all-results}
  The pion mass dependence of different quantities computed in the CQSM:
  the energy density in the center of the nucleon $\rho_E(0)$,
  the mean square radii $\la r_E^2\ra$ and $\la r_J^2\ra$ as defined in
  Eqs.~(\ref{Eq:def-energy-mean-square-radius},~\ref{Eq:def-spin-mean-square-radius}),
  the pressure $p(0)$ in the center of the nucleon,
  the zero of the pressure defined as $p(r_0)=0$,
  the constant $d_1$,
  the dipole masses of the form factors $M_2(t)$, $J(t)$ and $d_1(t)$
  as defined in Eq.~(\ref{Eq:dipol}),
  and the mean squared radius of the trace of the EMT (\ref{Eq:ff-of-EMT-trace-3}).
  In the chiral limit $J(t)$ and $d_1(t)$ have infinitely steep slopes at $t=0$.
  In these cases dipole fits do not provide useful approximations and are
  undefined (labelled by ``---'' in the Table).
  The proper time regularization can be applied, in principle, to any $m_\pi$
  \cite{Goeke:2005fs}.
  The Pauli-Villars regularization is applicable only for $m_\pi=0$
  \cite{Kubota:1999hx}. }
\vspace{0.2cm}
    \begin{ruledtabular}
    \begin{tabular}{cccccccccccc}
    \\
&
$m_\pi$     &
$\rho_E(0)$ &
$\la r_E^2\ra$  &
$\la r_J^2\ra$  &
$p(0)$      &
$r_0$       &
$d_1$       &
\multicolumn{3}{l}{dipole masses $M_{\rm dip}$ in GeV for}
& $\la r_F^2\ra$
\cr
&
${\rm MeV}$ &
${\rm GeV}/{\rm fm}^3$&
${\rm fm}^2$&
${\rm fm}^2$&
${\rm GeV}/{\rm fm}^3$&
${\rm fm}$ &
&
$M_2(t)$ &
$J(t)$   & 
$d_1(t)$ &
${\rm fm}^2$ \cr
    \\
    \hline
    \\
\multicolumn{5}{l}{{proper time regularization:}} &&&&&\cr
\cr
&0   & 1.54 & 0.79 & $\infty$ & 0.195 & 0.59 & -3.46 & 0.867 &  --- &  --- & 1.01 \\
&50  & 1.57 & 0.76 & 1.42     & 0.202 & 0.59 & -3.01 & 0.873 &  0.692  &  0.519  & 0.95 \\
&140 & 1.70 & 0.67 & 1.32     & 0.232 & 0.57 & -2.35 & 0.906 &  0.745  &  0.646  & 0.81\\
\\
\multicolumn{5}{l}{Pauli-Villars regularization:} &&&&&\cr
\cr
&0   & 0.75 & 0.86 & $\infty$ & 0.332 & 0.63 & -4.75 & 0.804 &  --- &  --- & 1.24  \\
\cr
\end{tabular}
\end{ruledtabular}
\end{table}
\endgroup
%

\newpage
\section{Conclusions}
\label{Sec-9:conclusions}

In this work we presented a study of the form factors of the EMT
in the large-$N_c$ limit in the framework of the CQSM.
We provided numerous checks of the theoretical consistency of the model results.
Among others, we demonstrated that the same model expressions for the form
factors can be derived from the EMT and from GPDs.

We computed the spatial density distributions, and mean square radii of the
operators of different components of the EMT and its trace.
Interesting results are that the energy density related to $\hat{T}^{00}$ in the
center of the nucleon is about $1.7\,{\rm GeV}^{-3}$, i.e.\ about 13 times
higher then the equilibrium density of nuclear matter.
The mean square radius of the operator $\hat{T}^{00}$
is about $0.67\,{\rm fm}^2$. For the mean square radius of the ``angular momentum
distribution'' related to the operator $\hat{T}^{0k}$ ($k=1,\,2,\,3$) we find
a much larger value $1.32\,{\rm fm}^2$.

We studied the spatial distribution of strong forces in the nucleon as described in
terms of the distributions of pressure and shear forces which are defined by the
spatial components $\hat{T}^{jk}$ ($j,\,k=1,\,2,\,3$) of the EMT. As a byproduct
of this study we learned how the soliton acquires stability in the CQSM ---
namely due to subtle balance of repulsive forces in the center due the ``quark core''
and attractive forces in the outskirts of the nucleon due the ``pion cloud''
which bounds the quarks in the model.

We observed a physically appealing connection between the criterion for the
stability of a particle, and the sign of the constant $d_1$, i.e.\
the form factor $d_1(t)$ at zero-momentum transfer.
Our observations imply that for a stable particle one always has $d_1<0$,
though we cannot prove this conjecture for the general case.
All available results for $d_1$ in literature are compatible with this observation.

We derived the leading non-analytic chiral contributions to the form factors
in the large $N_c$ limit, which agree with results from chiral perturbation theory
\cite{Chen:2001pv,Belitsky:2002jp,Diehl:2006ya} provided one takes into
account that the limits $N_c\to\infty$ and $m_\pi\to 0$ do not commute
\cite{Dashen:1993jt,Cohen:1992uy}.

We observed that the model results for the form factors of the EMT can,
for $m_\pi > 0$, be well approximated by dipole fits. The different form
factors have different $t$-dependences. For $J^Q(t)$ we obtain in the model
a dipole mass similar to that of the electromagnetic form factors of the proton.
The dipole mass of $M_2(t)$ is much larger than that, while that of $d_1(t)$
is substantially smaller. These results are of interest for the phenomenology
of hard exclusive reactions, and in particular for the task of extrapolating
$J^Q(t)$ to $t=0$ which is necessary to extract from data information on
what portion of the nucleon spin is due to quarks.

Our results yield in the chiral limit for the constant
$d_1= -(3.5-4.8)$ depending on the regularization and confirm sign and
--- within model accuracy --- magnitude of previous results
\cite{Petrov:1998kf,Kivel:2000fg}. We observe, however, also a
strong $m_\pi$-dependence of $d_1$ which is dominated by a
sizeable leading non-analytic (in the current quark mass)
contribution proportional to $m_\pi$. The latter is responsible
for reducing the absolute value of $d_1$ by about $30\%$ at the
physical point compared to its chiral limit value.

We estimated the mean square radius of the EMT trace operator to be about
$(1\,{\rm fm})^2$ which appears much larger than the corresponding mean square radius
of the traceless part of the EMT found to be $(0.3\,{\rm fm})^2$ \cite{Braun:1992jp},
and noticed that the instanton vacuum model provides a possible explanation for that.

A study of the $m_\pi$-dependence of the EMT form factors in the model
at larger values of $m_\pi$ in the spirit of Ref.~\cite{Goeke:2005fs}
and a detailed comparison of the model results to lattice QCD data
\cite{Mathur:1999uf,Gadiyak:2001fe,Hagler:2003jd,Gockeler:2003jf,Negele:2004iu,Schroers:2007qf}
will be presented elsewhere \cite{accompanying-paper}.

\vspace{1cm}
\noindent
{\bf Acknowledgements} \hspace{0.2cm}
We thank Pavel Pobylitsa for fruitful discussions and valuable
comments. This research is part of the EU integrated infrastructure initiative
hadron physics project under contract number RII3-CT-2004-506078,
and partially supported by the Graduierten-Kolleg Bochum-Dortmund
and Verbundforschung of BMBF.
A.~S.\  acknowledges support from GRICES and DAAD.

\vspace{1cm}
\noindent
{\bf Note added} \hspace{0.2cm}
After this work was completed the work \cite{Wakamatsu:2007uc} appeared
where in particular the constant $d_1$ was studied. For $d_1^{u+d}$
similar results were obtained using somehow different model parameters.
Interesting is the estimate for the flavour combination $d_1^{u-d}$ which
was found rather small and confirms the large-$N_c$ prediction
(\ref{Eq:D-term-in-large-Nc}).

\newpage

\appendix

\section{\boldmath Alternative definition of form factors}
\label{App:Alternative-definition}

The following alternative definition of form factors of the EMT is commonly
used in literature, see e.g.\ \cite{Ji:1996ek},
\ba
    \la p^\prime| \hat T_{\mu\nu}^{Q,G}(0) |p\rangle
    &=& \bar u(p^\prime)\biggl[
    A^{Q,G}(t)\,\frac{\gamma_\mu P_\nu+\gamma_\nu P_\mu}{2}+
    B^{Q,G}(t)\,\frac{i(P_{\mu}\sigma_{\nu\rho}+P_{\nu}\sigma_{\mu\rho})
    \Delta^\rho}{4M_N}
    \nonumber \\
    &+& C^{Q,G}(t)\,\frac{\Delta_\mu\Delta_\nu-g_{\mu\nu}\Delta^2}{M_N}
    \pm \bar c(t)g_{\mu\nu} \biggr]u(p)\, .
    \label{Eq-app:ff-of-EMT-alternative} \ea
By means of the Gordon identity
$2M_N\bar u^\prime\gamma^\alpha u=\bar u^\prime(i\sigma^{\alpha\kappa}\Delta_\kappa+2P^\alpha)u$
Eq.(\ref{Eq-app:ff-of-EMT-alternative}) can be rewritten as
Eq.(\ref{Eq:ff-of-EMT}) with
\ba
    A^{Q,G}(t)            &=& M_2^{Q,G}(t) \phantom{\frac11}\nonumber\\
    A^{Q,G}(t)+B^{Q,G}(t) &=& 2\,J^{Q,G}(t)\nonumber\\
    C^{Q,G}(t)            &=& \frac15\,d_1^{Q,G}(t)\,.
    \label{Eq-app:ff-of-EMT-alternative2}\ea
The constraints (\ref{Eq:M2-J-d1}) translate in this language into $A^Q(0)+A^G(0)=1$
and $B^Q(0)+B^G(0)=0$. The latter constraint means that the total nucleon
"gravitomagnetic moment" vanishes.

In models, in which the only dynamical degrees of freedom are effective quark
degrees of freedom, the constraint $B^Q(0)=0$ must hold. Such is the situation in
the CQSM where consequently this constraint is satisfied \cite{Ossmann:2004bp}.

Interestingly, it was argued \cite{Teryaev:1998iw} that also in QCD the quark
and gluon gravitomagnetic moments of the nucleon could vanish separately,
i.e.\ $B^Q(0)=0$ and $B^G(0)=0$.
That would imply that $M_2^Q(0)=2J^Q(0)$ and $M_2^G(0)=2J^G(0)$ at any scale, and
not only in the asymptotic limit of a large renormalization scale \cite{Ji:1996ek},
see \cite{Teryaev:1998iw} for details.

\section{\boldmath General relations from the conservation of EMT}
\label{App:general-relations-from-EMT-conservation}

Here we collect some worthwhile noticing relations for $p(r)$ and $s(r)$
which can be derived from the differential equation (\ref{Eq:p(r)+s(r)}) ---
i.e.\  which follow from the conservation of the EMT.
\begin{itemize}
\item   For {\sl any} $s(r)$ one obtains from (\ref{Eq:p(r)+s(r)}) a $p(r)$
    which {\sl automatically} satisfies the stability condition
    (\ref{Eq:stability}).
    Therefore, when computing these quantities for example in a model,
    the computation of the pressure is more important and reliable in the
    sense that the result can be cross-checked by the stability condition
    (\ref{Eq:stability}).

\item   The pressure at the origin is connected to $s(r)$ by the following
    integral relation
    \be\label{Eq:p(r)+s(r)-consequence-01}
       p(0) = 2\int\limits_0^\infty\di r\;\frac{s(r)}{r}\;.
    \ee

\item   Assume that $p(r)$ and $s(r)$ vanish at large $r$ faster than any
    power of $r$ to justify below integration by parts.
    (In Sec.~\ref{Sec-7:pressure+shear} we have seen that this is always
    the case with the exception of the chiral limit.) Then the following
    relations hold between the ``Mellin moments'' of $s(r)$ and $p(r)$
    \be\label{Eq:s(r)-p(r)-integral-relations}
       \int\limits_0^\infty\di r\;r^N\, s(r) = -\,\frac{3(N+1)}{2(N-2)}
       \int\limits_0^\infty\di r\;r^N\, p(r)\;\;\;\;\mbox{for}\;\;\; N > -1,
    \ee
    which are valid also for non-integer values of $N$.
    Eq.~(\ref{Eq:d1-from-s(r)-and-p(r)}) quoted in
    Sec.~\ref{Sec-2:FF-of-EMT-in-general} is just a special case of
    (\ref{Eq:s(r)-p(r)-integral-relations}).

\item   The case $N=2$ in (\ref{Eq:s(r)-p(r)-integral-relations}) has to be
    treated with care. Taking the limit $N\to 2$ on the right-hand-side
    of (\ref{Eq:s(r)-p(r)-integral-relations}) yields --- upon use of
    (\ref{Eq:stability}) --- the following result
    \be\label{Eq:s(r)-p(r)-integral-relations-a}
        \int\limits_0^\infty\di r\;r^2\, s(r) = -\,\frac{9}{2}
        \int\limits_0^\infty\di r\;r^2\,{\rm log}\,r \; p(r)\;.\ee
    One may worry in which units the ``$r$'' in ``$\log r$'' should be
    provided in (\ref{Eq:s(r)-p(r)-integral-relations-a}). However, this
    is immaterial due to the stability condition (\ref{Eq:stability}).

\item   $p(r)$ and $r^3s(r)$ exhibit extrema at the same $r$.

\end{itemize}

\section{\boldmath Conservation of EMT in the CQSM}
\label{App:conservation-of-EMT-in-model}

In this appendix we demonstrate explicitly that
$\la N^\prime|\partial_\mu\hat{T}_{\rm eff}^{\mu\nu}|N\ra=0$.
For that we directly evaluate in the model matrix elements of
the operator $\partial_\mu\hat{T}_{\rm eff}^{\mu\nu}$.
For $\partial_0\hat{T}_{\rm eff}^{0\nu}$ a derivation analog to that
yielding (\ref{Eq:T00-in-model}) and (\ref{Eq:T0k-in-model}) yields
immediately
\be\label{App-conservation-01}
    \la N^\prime|\partial_0\hat{T}_{\rm eff}^{0\nu}|N\ra=0\;.
\ee
Hereby one has to consider the cases $\nu=0$ and $\nu=k$ separately,
since the respective model expressions arise from different orders in
the $1/N_c$ expansion.

A derivation analog to that yielding (\ref{Eq:Tik-in-model}) gives for matrix
elements of the operator $\partial_i\hat{T}_{\rm eff}^{ik}$ the following result
\ba
    \la N^\prime|\partial_i\hat{T}_{\rm eff}^{ik}|N\ra
    \!\! &=& \!\!
    \delta_{S_3^\prime S_3^{\phantom{\prime}}}
    \frac{M_NN_c}{2} \!\sum_{n,\rm occ}
    \int\!\di^3{\bf x} \,e^{i{\bf\Delta x}}\,i\nabla^i
    \biggl[
     \phi_n^\ast\gamma^0\gamma^i(\nabla^k\phi_n)
    +\phi_n^\ast\gamma^0\gamma^k(\nabla^i\phi_n)
    -(\nabla^k\phi_n^\ast)\gamma^0\gamma^i\phi_n
    -(\nabla^i\phi_n^\ast)\gamma^0\gamma^k\phi_n\biggr] \nonumber\\
    &=& \! \delta_{S_3^\prime S_3^{\phantom{\prime}}}
    \frac{M_NN_c}{2} \sum_{n,\rm occ}
    \int\!\di^3{\bf x} \,e^{i{\bf\Delta x}} (K_1+K_2 )
    \label{App-conservation-02}
\ea
where
\ba
    K_1 &=&
    -\phi_n^\ast [\nabla^k(-i\gamma^0\gamma^i\nabla^i)\phi_n)]
    +[-i\gamma^0\gamma^i\nabla^i\phi_n]^\ast [\nabla^k\phi_n]
    +(\nabla^k\phi_n^\ast)(-i\gamma^0\gamma^i\nabla^i\phi_n)
    -[\nabla^k(-i\gamma^0\gamma^i\phi_n)^\ast]\phi_n \nonumber\\
    K_2 &=&
    +i\phi_n^\ast\gamma^0\gamma^k({\bf\nabla}^2\phi_n)
    -i(\nabla^2\phi_n^\ast)\gamma^0\gamma^k\phi_n \;.
   \label{App-conservation-03} \ea
From the single quark equations of motion (\ref{Hamiltonian}) one obtains the
identities $(-i\gamma^0\gamma^i\nabla^i)\phi_n = (E_n-M\gamma^0U^{\gamma_5})\phi_n$
and $(-\nabla^2\phi_n)=(E_n^2-M^2-iM\gamma^j(\nabla^jU^{\gamma_5}))\phi_n$ which
allow respectively to rewrite $K_1$ and $K_2$ as follows
\ba\label{App-conservation-04}
    K_1 = -K_2 = 2 M\phi_n^\ast\gamma^0(\nabla^k U^{\gamma_5})\phi_n \;.
\ea
Thus, $\la N^\prime|\partial_i\hat{T}_{\rm eff}^{ik}|N\ra=0$ in
Eq.~(\ref{App-conservation-02}). This proves that the form factor $\bar c(t)$
in Eq.~(\ref{Eq:ff-of-EMT}) vanishes in the model. This is consistent since,
due to the  absence of gluons, in the CQSM the quark part of the EMT must
be conserved by itself.

\section{\boldmath Regularization}
\label{App:regularization}

The proper time regularized model expressions for the continuum contributions
to the form factors (\ref{Eq:M2-d1-model},~\ref{Eq:d1-model},~\ref{Eq:J-model})
read
\ba
&&  M_2(t)_{\rm cont}-\frac{t}{5M_N^2}d_1(t)_{\rm cont}
    =
    \frac{N_c}{M_N}\sum\limits_{n,\,\rm all}R_1(E_n,\Lambda)
    \la n|e^{i{\bf \Delta}\hat{\bf x}}|n\ra\;,
    \label{Eq:M2-d1-model-reg} \\
&&  d_{1}(t)_{\rm cont}
    =
    \frac{5M_NN_c}{4t}
    \sum_{n,\rm all}R_2(E_n,\Lambda)\langle n|
    \biggl\{\gamma^0\mbox{\boldmath$\gamma$}\hat{\bf p},
    \,e^{i{\bf \Delta}\hat{\bf x}}\biggr\} |n\rangle\;,
    \label{Eq:d1-model-reg} \\
&&  J(t)_{\rm cont}
    =
    \frac{iN_c\,\varepsilon^{klm}\Delta^k}{8I t}
    \doublesum{n,j\,\rm all}{n\neq j} R_3(E_n,E_j,\Lambda)
    \langle n|\tau^l|j\rangle
    \langle j|\biggl(
    \biggl\{ e^{i{\bf \Delta}\hat{\bf x}},\hat{p}^m\biggr\}
    +(E_n+E_j)e^{i{\bf \Delta}\hat{\bf x}}\gamma^0\gamma^m\biggr)|n\rangle
    \label{Eq:J-model-reg} \ea
with the regulator functions $R_i$ defined as follows
\ba
&&  R_1(\omega,\Lambda)
    =\frac{1}{4\sqrt{\pi}}\int\limits_{1/\Lambda^2}^\infty
    \frac{\di \alpha }{\alpha^{3/2}}\;\exp(-\alpha \omega^2)\nonumber\\
&&  R_2(\omega,\Lambda) = -\,\frac12\;\frac{\partial\;}{\partial\omega}
    R_1(\omega,\Lambda)\nonumber\\
&&  R_3(\omega_1,\omega_2,\Lambda)
    = \frac{1}{4\sqrt{\pi}}\int\limits_{1/\Lambda^2}^\infty\!\di\alpha\biggl[
        \frac{\exp(-\alpha \omega_1^2)-\exp(-\alpha \omega_2^2)}
         {\alpha^{3/2}(\omega_1^2-\omega_2^2)}
       -\frac{\omega_1\exp(-\alpha \omega_1^2)+\omega_2\exp(-\alpha \omega_2^2)}
         {\alpha^{1/2}(\omega_1+\omega_2)}\biggr]\label{App-reg-02}\ea
That $d_1(t)$ is regularized ``differently'' in Eqs.~(\ref{Eq:M2-d1-model-reg})
and (\ref{Eq:d1-model-reg}) is a peculiarity of the proper time regularization.
In the Pauli-Villars regularization all quantities $F=\{M_N,I,M_2(t),J(t),d_1(t)\}$
are regularized in the same way as
\be\label{App-reg-07}
    F_{\rm reg} = F(M)-\frac{M^2\,}{M_{PV}^2} \;F(M_{PV})\;,
\ee
where it is understood that the corresponding model expressions are first evaluated
with the Hamiltonian (\ref{Hamiltonian}) and with the Hamiltonian (\ref{Hamiltonian})
where $M$ replaced by $M_{PV}$, and then finally subtracted according to the
prescription (\ref{App-reg-07}).

Notice that for the constraints (\ref{App-constraints-01}) and
(\ref{Eq:J-model-comp-02}) to be satisfied by the numerical results
it is of crucial importance that $M_2(t)$ and $M_N$ as well as
$J(t)$ and $I$ are regularized in the same way.
This is the case for both regularizations.
In order to make this apparent also for the proper-time regularization
we recall that in this method the regularized  model expressions for continuum
contributions to the nucleon mass (\ref{Eq:soliton-energy}) and the moment of
inertia (\ref{Eq:mom-inertia}) are given by
\ba
&&  M_{N,\,\rm cont} \equiv N_c\sum_{E_n<0}(E_n-E_{n_0})|_{\rm reg}
    = N_c \sum_{n,\,\rm all}\biggl(R_1(E_n,\Lambda) -R_1(E_{n_0},\Lambda)\biggr)
    \nonumber\\
&&  I_{\rm cont} \equiv \frac{N_c}{6}\doublesum{E_m>0}{E_n < 0}
        \frac{\la n|\tau^a|m\ra\,\la m|\tau^a|n\ra}{E_m-E_n}\biggl|_{\rm reg}
    = \frac{N_c}{6}\doublesum{n,m\,\rm all}{m\neq n}
    \la n|\tau^a|m\ra\,\la m|\tau^a|n\ra R_3(E_n,E_m,\Lambda)\;.
    \label{App-reg-01}\ea
Hereby $\Lambda$ or $M_{PV}$ are fixed to reproduce the physical value of the
pion decay constant $f_\pi=93\,{\rm MeV}$ \cite{Christov:1995vm}.
For $M=350\,{\rm MeV}$ one has $\Lambda=649\,{\rm MeV}$ in the chiral limit,
and $\Lambda=643\,{\rm MeV}$ for $m_\pi=140\,{\rm MeV}$. The proper time
method can, in principle, be applied to any $m_\pi$ \cite{Goeke:2005fs}.
The Pauli-Villars method can be applied unambiguously in the
chiral limit, with $M_{PV}=556\,{\rm MeV}$ reproducing the
experimental value of $f_\pi$. However, the method meets difficulties
in the case of a non-zero current quark mass $m$ \cite{Kubota:1999hx}.
Notice that $\Lambda$ and $M_{PV}$ are of
${\cal O}(\rho^{-1})\approx 600\,{\rm MeV}$.

\section{\boldmath Model expressions for $d_1(0)$ and $d_1^{\,\prime}(0)$ from the EMT}
\label{App:d1(0)-and-d1prime(0)}

One consequence of the conservation of the EMT,
    see Sec.~\ref{Sec-2:FF-of-EMT-in-general} and
    Appendices~\ref{App:general-relations-from-EMT-conservation} and
    \ref{App:conservation-of-EMT-in-model},
is that there are two different expressions for $d_1(t)$. In one $d_1(t)$ is
related to $p(r)$ and in the other to $s(r)$. Both are, of course, equivalent.
However, this is not obvious from the explicit model expressions.

Here we derive model expressions for $d_1(0)$ and $d_1^{\,\prime}(0)$
from the EMT in terms of $p(r)$ and $s(r)$ which will be useful below for the
explicit demonstration that the expressions for $d_1(t)$ obtained from the EMT
and GPDs are equivalent.

The model expression for the spatial components
of the static energy-momentum tensor
(\ref{Def:static-EMT}) reads
\be\label{App-stability-00}
    T^{kl}({\bf r}) = \frac{N_c}{4i}\sum_{n,\,\rm occ}\phi_n^\ast({\bf r})
    \biggl(-\gamma^0\gamma^k\overleftarrow{\nabla}^l
           +\gamma^0\gamma^k\overrightarrow{\nabla}^l
           + (k\leftrightarrow l)\biggr)\phi_n({\bf r})\;.
\ee
For the model expressions for the form factor
$d_1(t)$ and its derivative at $t=0$ we obtain
\ba\label{Eq:d1(0)}
  &&    d_1(0)
    =
    \frac{5N_cM_N}{4}\,\sum_{n,\,\rm occ}
    \la n|\,\biggl\{\gamma^0{\bgam}\hat{\bf p},\,\hat{\bf r}^2\biggr\}|n\ra
    =
    -\frac{N_cM_N}{4}\, \sum_{n,\,\rm occ}
    \la n|\,\biggl\{\gamma^0\gamma^k\hat{p}^l,\;\left(
    \hat{r}^k\hat{r}^l-\frac{\hat{\bf r}^2}3\,\delta^{kl}\right)\biggr\}|n\ra
    \;,\\
\label{Eq:d1prime(0)}
  &&    d_1^{\,\prime}(0)
    =
        \frac{N_cM_N}{96}\sum_{n,\rm occ}\langle n|\biggl\{
    \gamma^0\mbox{\boldmath$\gamma$}\hat{\bf p},\,\hat{\bf r}^4\biggr\} |n\rangle
    =
    -\frac{N_cM_N}{56}\sum_{n,\,\rm occ}\la n|\biggr\{\gamma^0\gamma^k\hat{p}^l,\,
    \left(\hat{r}^k\hat{r}^l-\frac{\hat{\bf r}^2}3\,\delta^{kl}\right)
    \hat{\bf r}^2\,\biggr\}|n\ra
    \;. \ea
The first relations in (\ref{Eq:d1(0)}) and (\ref{Eq:d1prime(0)}) are more practical
for a numerical evaluation. They follow for example from taking the limit
$\Delta^i\to 0$ in (\ref{Eq:d1-model}) and making use of (\ref{Eq:stability})
and hedgehog symmetry. Notice, that (\ref{Eq:d1(0)}) can alternatively be derived
from (\ref{Eq:d1-from-s(r)-and-p(r)}) and (\ref{Eq:density-pressure}) --- which is
a cross check for the intermediate model expressions.

In order to find the equivalent second relations in (\ref{Eq:d1(0)}) and
(\ref{Eq:d1prime(0)}) we inspect Eq.~(\ref{Eq:ff-J}) for small momentum transfers
using for $T^{ij}({\bf r})$ the model expression in (\ref{App-stability-00}).
This yields
\be\label{Eq:d1prime(0)-prepare}
        d_1(0) + t\;\frac{7}{3}\, d_1^{\,\prime}(0) + {\cal O}(t^2)
    =
    -\frac{N_cM_N}{4}\sum_{n,\,\rm occ}\la n|\biggr\{\gamma^0\gamma^i\hat{p}^j,\,
    \left(\hat{r}^i\hat{r}^j-\frac{\hat{\bf r}^2}3\,\delta^{ij}\right)
    \biggl(1+\frac{\hat{\bf r}^2t}{6}+{\cal O}(t^2)\biggr)\,\biggr\}|n\ra\;,
\ee
from which we read off the desired results. The second relation in
(\ref{Eq:d1prime(0)}) follows also immediately from (\ref{Eq:M2-J-d1})
and (\ref{App-stability-00}).

Comparing (\ref{App-stability-00}) with the general relations for $p(r)$
and $s(r)$ with $T^{ij}(r)$,
\be\label{App-stability-04}
    p(r) = \frac13\;T^{ij}(r)\;\delta^{ij}\;,\;\;\;\;
    s(r) = \frac32\;T^{ij}(r)\,\biggl(\frac{r^ir^j}{r^2}-\frac13\delta^{ij}
    \biggr),
\ee
we recognize that (\ref{Eq:d1(0)}) is just Eq.~(\ref{Eq:d1-from-s(r)-and-p(r)})
in the model, while  (\ref{Eq:d1prime(0)}) is the model version of
Eq.~(\ref{Eq:d1-slope-at-t=0}).

\newpage
\section{\boldmath Polynomiality of $(H^u+H^d)(x,\xi,t)$ at $t\neq 0$}
\label{App:polynomiality}

The expression for $(H^u+H^d)(x,\xi,t)$, which in the SU(2) version of the CQSM
already exhausts the sum over quark flavours in Eq.~(\ref{Eq:EMT-and-GPD-H}),
was derived, evaluated and discussed in \cite{Petrov:1998kf}.
With the light-like vector $n$ in Eq.~(\ref{Def:H-and-E}) chosen to
be $n^\mu\propto(1,0,0,-1)$ the model expression is given by
\be
        (H^u+H^d)(x,\xi,t) = M_N N_c\int\frac{\di z^0}{2\pi}
        \sum\limits_{n,\rm occ} e^{iz^0(xM_N-E_n)}
        \la n|\,(1+\gamma^0\gamma^3)\exp(-i\fracS{z^0}{2}\hat{p}^3)
        \exp(i\bDelta\hat{\bf x})\exp(-i\fracS{z^0}{2}\hat{p}^3)\,|n\ra
        \;.\label{App1-Eq:Hu+Hd-model} \ee
The study of Ref.~\cite{Petrov:1998kf} was supplemented in \cite{Schweitzer:2002nm} by
the explicit proof that $(H^u+H^d)(x,\xi,t)$ in the model satisfies the polynomiality
condition (\ref{Def:polynom-Hq}) at $t=0$. In this Appendix we generalize the
proof of Ref.~\cite{Schweitzer:2002nm} to $t\neq 0$.

The reason why the proof of \cite{Schweitzer:2002nm} was restricted to the case $t=0$
is connected to the fact that the information on $\xi$ and $t$ on the RHS of
Eq.~(\ref{App1-Eq:Hu+Hd-model}) is {\sl implicitly} encoded in the 3-vector
${\bf\Delta}$. In the large-$N_c$ kinematics $\Delta^3 = -2\xi M_N$ and
$t=-{\bf\Delta}^2$. By keeping $\xi\neq 0$ and continuing analytically the moments
of $(H^u+H^d)(x,\xi,t)$ to the point $t=0$ one obtains model expressions depending
on $\xi$ only. That simplifies the situation considerably \cite{Schweitzer:2002nm}.

However, the proof of \cite{Schweitzer:2002nm} can be generalized to $t\neq 0$
by using the following remarkable identity
\be\label{App1-Eq:identity}
    \exp(i\bDelta{\bf x})=\sum\limits_{l=0}^\infty i^l(2l+1)
    j_l(r\sqrt{-t})\,P_l(\cos{\theta})\,
    P_l\biggl(-\frac{2\xi M_N}{\sqrt{-t}}\biggr)\;,
\ee
where $r=|{\bf x}|$ and $\cos{\theta} = {x}^3/|{\bf x}|$ and
$j_l(z)$ and $P_l(z)$ denote respectively Bessel functions and Legendre polynomials.
As a byproduct we remark that by inserting the identity (\ref{App1-Eq:identity}) in
(\ref{App1-Eq:Hu+Hd-model}) we see {\sl explicitly} that $(H^u+H^d)(x,\xi,t)$ is a
function only of $\xi$ and $t$ (and $x$, of course) but not of the 3-vector $\bDelta$.

The identity (\ref{App1-Eq:identity}) can be derived as follows.
We rewrite the 3-vector ${\bf\Delta}$ as
${\bf\Delta}=(\sin\delta\cos\alpha,\sin\delta\sin\alpha,\cos\delta)\sqrt{-t}$
where $\cos\delta=-2\xi M_N/\sqrt{-t}$ and $\alpha\in[0;2\pi]$ describes the
orientation of ${\bf\Delta}$ in the plane perpendicular to the spatial direction
of the light-cone. Then, by using spherical coordinates for ${\bf x}$, we obtain
\be\label{Eq:derive-identity-1}
        \exp(i\bDelta{\bf x}) = \sum\limits_{l=0}^\infty i^l(2l+1)j_l(r\sqrt{-t})\,
    P_l(\cos{\theta}\cos\delta+\sin\theta\sin\delta\cos(\alpha-\phi))\;.
\ee
Next we rewrite the Legendre polynomials using the addition theorem as
\be\label{Eq:derive-identity-2}
    P_l(\cos{\theta}\cos\delta+\sin\theta\sin\delta\cos(\alpha-\phi))
    = P_l(\cos\theta)P_l(\cos\delta)
    + 2 \sum\limits_{j=1}^l\,\frac{(l-j)!}{(l+j)!}\,
    P_l^j(\cos\theta)P_l^j(\cos\delta)\,\cos(l(\alpha-\phi)) \;,
\ee
The physical process described by the GPD does not depend on the orientation of
${\bf\Delta}$ in the transverse plane. This is reflected by the fact that the
matrix elements in (\ref{App1-Eq:Hu+Hd-model}) are independent of $\alpha$.
(Changes of $\alpha$ can be compensated by appropriate hedgehog rotations.)
Thus, we can eliminate the artificial dependence on $\alpha$ --- for example
by averaging the matrix elements in (\ref{App1-Eq:Hu+Hd-model}) over $\alpha$.
The latter is equivalent to taking the average over $\alpha$ in
(\ref{Eq:derive-identity-2}). Inserting the result into (\ref{Eq:derive-identity-1})
gives finally the identity (\ref{App1-Eq:identity}).

Notice that for $\epsilon \to 0$ the product
$(2l+1)\,j_l(A\epsilon)\,P_l(B/\epsilon)\to(AB)^l/l!+{\cal O}({\epsilon^2})$.
Thus, taking the limit $t\to 0$ while keeping $\xi\neq 0$ in (\ref{App1-Eq:identity})
yields
\be\label{App1-Eq:identity-at-t=0}
    \triplelim{\rm analytical}{\rm continuation}{t\to 0,\;\xi\neq 0}
    \exp(i\bDelta{\bf x})
    = \sum\limits_{l=0}^\infty \frac{(-i 2\xi M_N|{\bf x}|)^l\!}{l!}\;
    P_l(\cos{\theta})
    \;,\label{App:an-cont-II-5} \ee
reproducing Eq.~(20) of Ref.~\cite{Schweitzer:2002nm} which was derived
independently in a different way.

The proof of polynomiality at $t=0$ given in Ref.~\cite{Schweitzer:2002nm} is
generalized to any $t\neq 0$ by replacing Eq.~(20) in \cite{Schweitzer:2002nm}
by our more general identity (\ref{App1-Eq:identity}) and repeating literally
the steps done in Eqs.~(21-28) in \cite{Schweitzer:2002nm}. This yields
\ba
    \int\!\di x\;x^{m-1}\,(H^u+H^d)(x,\xi,t)
    &=&
    \frac{N_c}{M_N^{m-1}} \sum\limits_{n,\rm occ}
    \sum\limits_{k=0}^{m-1}\left(\matrix{m-1\cr k}\right)
    \frac{E_n^{m-1-k}}{2^k}
    \sum\limits_{l=0,2,4,\dots}^{k+1}\,
    i^l(2l+1)P_l\biggl(\frac{2\xi M_N}{\sqrt{-t}}\biggr)
    \nonumber\\
    && \times \sum\limits_{j=0}^k\left(\matrix{k\cr j}\right)
    \la n|(\gamma^0\gamma^3)^k\,(\hat{p}^3)^{j}
    j_l(|\hat{\bf x}|\sqrt{-t})\,P_{l}(\cos\hat{\theta})
    \,(\hat{p}^3)^{k-j}|n\ra \;,
    \label{App1-Eq:polynomiality}
\ea
and explicitly demonstrates that the model expression
(\ref{App1-Eq:Hu+Hd-model}) does satisfy the polynomiality property for
any $t\neq 0$. Note that this includes also positive $t$ allowing to study
in principle the region of time-like momentum transfers, where GPDs ``become''
nucleon-antinucleon distribution amplitudes \cite{Diehl:1998dk,Freund:2002cq}.

\section{\boldmath $M_2(t)$ and $d_1(t)$ from GPDs}
\label{App:FFs-from-EMT-and-GPD}

In this and the following Appendix we introduce labels to distinguish the model
expressions for form factors derived from GPDs and EMT --- with the aim to prove
finally that they are equivalent.
From Eq.~(\ref{App1-Eq:polynomiality}) we read off the model expressions for the
second Mellin moment of $(H^u+H^d)(x,\xi,t)$, see Eq.~(\ref{Eq:EMT-and-GPD-H}),
\ba
        M_2(t)|_{\rm GPD}
        &=& \frac{N_c}{M_N} \sum\limits_{n,\rm occ}
            \la n| E_n j_0(|\hat{\bf x}|\sqrt{-t})|n\ra
    +\frac{N_c}{2M_N} \sum\limits_{n,\rm occ}\la n|\biggl[
            \{ \gamma^0\gamma^3\hat{p}^3,j_0(|\hat{\bf x}|\sqrt{-t})
         +  \frac52\,j_2(|\hat{\bf x}|\sqrt{-t})\,P_2(\cos\hat{\theta})\}\biggr]\,
            |n\ra \;,
    \nonumber\\
\label{App2-Eq:M2-d1-from-GPD}
        d_1(t)|_{\rm GPD}
        &=& \frac{75\,N_c M_N}{4t} \sum\limits_{n,\rm occ} \la n|\,
            \{ \gamma^0\gamma^3\hat{p}^3,j_2(|\hat{\bf x}|\sqrt{-t})\,
            P_2(\cos\hat{\theta})\}\,|n\ra \;.\ea
By comparing the expressions for $M_2(t)|_{\rm GPD}$ and $d_1(t)|_{\rm GPD}$
in (\ref{App2-Eq:M2-d1-from-GPD}),
and by exploring hedgehog symmetry we observe
\ba
    \frac{5N_c}{4M_N}\sum\limits_{n,\rm occ}\la n|\{ \gamma^0\gamma^3\hat{p}^3,
    \,j_2(|\hat{\bf x}|\sqrt{-t})\,P_2(\cos\hat{\theta})\}|n\ra
    = \frac{t}{15M_N^2}\,d_1(t)_{\rm GPD}
    \nonumber\\
    \frac{N_c}{M_N} \sum\limits_{n,\rm occ}
        \la n| E_n j_0(|\hat{\bf x}|\sqrt{-t})|n\ra
    = \frac{N_c}{M_N} \sum\limits_{n,\rm occ}
          \la n| E_n e^{i{\bf\Delta\hat{x}}}|n\ra
    \equiv [M_{2}(t)-\frac{t}{5M_{N}^{2}}d_{1}(t)]_{\rm EMT}
    \nonumber\\
    \frac{N_c}{2M_N} \sum\limits_{n,\rm occ}\la n|
        \{ \gamma^0\gamma^3\hat{p}^3,j_0(|\hat{\bf x}|\sqrt{-t})\}|n\ra
    = \frac{N_c}{6M_N} \sum\limits_{n,\rm occ}\la n|
        \{ \gamma^0{\bf\gamma}\hat{\bf p},e^{i{\bf\Delta\hat{x}}}\}|n\ra
    \equiv \frac{2t}{15M_N^2}\,d_1(t)_{\rm EMT}\;.
\ea
Thus,  Eq.~(\ref{App2-Eq:M2-d1-from-GPD}) states that
\be\label{App2-Eq:M2-from-GPD-II}
       [M_2(t)-\frac{t}{15M_N^2}\,d_1(t)]_{\rm GPD}
    =  [M_2(t)-\frac{t}{15M_N^2}\,d_1(t)]_{\rm EMT}\;.\ee
This means that the model expressions from GPDs and from EMT are equivalent
for this particular linear combination of $M_2(t)$ and $d_1(t)$.
In order to prove that this is true also for the separate form factors
we have show explicitly that e.g.\ one can derive the same model expression
for $d_1(t)_{\rm GPD}$ from the EMT. One way to do this is to demonstrate that
$d_1(t)_{\rm GPD}$ satisfies the differential equation (\ref{Eq:ff-d1})
{\sl derived from the EMT} with appropriate boundary conditions.

For that let us first remove the preference of the 3-axis from the expression for
$d_1(t)|_{\rm GPD}$ in (\ref{App2-Eq:M2-d1-from-GPD}) which is due to arbitrarily
choosing the light-like vector $n^\mu\propto(1,0,0,-1)$, see above. We remove this
arbitrariness by averaging the expression (\ref{App2-Eq:M2-d1-from-GPD}) over
directions. This yields
\ba\label{Eq:App-av}
        \biggl\la\biggl\{\,\gamma^0\gamma^3\hat{p}^3,\;
        P_{2}(\cos\hat\theta)j_2(|\hat{\bf x}|\sqrt{-t})\biggr\}\biggr\ra_{\!\rm av.}
        &=&
        \frac{1}{5}\,\biggl\{\gamma^0\gamma^k\hat{p}^l,\;
    \biggl(\hat{x}^l\hat{x}^k-\frac{1}{3}\,|\hat{\bf x}|^2\delta^{kl}\biggr)
    \,\frac{j_2(|\hat{\bf x}|\sqrt{-t})}{|\hat{\bf x}|^2}\biggr\}\;
\ea
and using the identity
\be
    \biggl(1+\frac{4t}{3}\;\frac{\di\,}{\di t}
    +\frac{4t^2}{15}\;\frac{\di^2\;}{\di t^2}\biggr)
    \frac{j_2(|\hat{\bf x}|\sqrt{-t})}{|\hat{\bf x}|^2(-t)} =
    \frac{j_0(|\hat{\bf x}|\sqrt{-t})}{15}
\ee
we see that $d_1(t)_{\rm GPD}$ satisfies the differential equation (\ref{Eq:ff-d1})
\ba
        [d_1(t)+\frac{4t}{3}\, d_1^{\,\prime}(t)
    +\frac{4t^2}{15}\, d_1^{\,\prime\prime}(t)]_{\rm GPD}
        &=&
    -\,\frac{N_c M_N}{4} \sum\limits_{n,\rm occ} \la n|\,
        \biggl\{\gamma^0\gamma^k\hat{p}^l,\;
    \biggl(\hat{x}^l\hat{x}^k-\frac{1}{3}\,|\hat{\bf x}|^2\delta^{kl}\biggr)
    \,j_0(|\hat{\bf x}|\sqrt{-t})\biggr\}\,|n\ra
    \nonumber\\
    &\equiv&
    -\frac{M_N}{2}\, \int\di^3{\bf r}\,e^{-i{\bf r}\bDelta}\,
        T_{ij}^Q({\bf r})\,\left(r^i r^j-\frac{{\bf r}^2}3\,\delta^{ij}\right)\;.
\ea
Next, using (\ref{Eq:App-av}) and expanding the expression for $d_1(t)_{\rm GPD}$
in (\ref{App2-Eq:M2-d1-from-GPD}) we obtain
\ba
        d_1(0)|_{\rm GPD}
        &=&
    -\,\frac{N_c M_N}{4} \sum\limits_{n,\rm occ} \la n|\,
    \biggl\{\gamma^0\gamma^k\hat{p}^l,\;
    \biggl(\hat{x}^l\hat{x}^k-\frac{1}{3}\,|\hat{\bf x}|^2\delta^{kl}\biggr)
    \biggr\}\,|n\ra \nonumber\\
        d_1^{\,\prime}(0)|_{\rm GPD}
        &=& -\,\frac{N_c M_N}{56} \sum\limits_{n,\rm occ} \la n|\,
    \biggl\{\gamma^0\gamma^k\hat{p}^l,\;
    \biggl(\hat{x}^l\hat{x}^k-\frac{1}{3}\,|\hat{\bf x}|^2\delta^{kl}\biggr)
    |\hat{\bf x}|^2\biggr\}\,
    |n\ra \;,\ea
which coincides with the expressions obtained from the EMT in Eqs.~(\ref{Eq:d1(0)})
and (\ref{Eq:d1prime(0)}). This, completes the proof that one obtains the same model
expressions for the form factors $d_1(t)$ and $M_2(t)$ from GPDs and from the EMT.

\section{\boldmath $J(t)$ from GPDs}
\label{App:FFs-from-EMT-and-GPD-2}

In this Appendix we show that the model expression for $J(t)_{\rm EMT}$
obtained from the EMT in Eq.~(\ref{Eq:J-model}) coincides with the model
expression for $J(t)_{\rm GPD}$ which results from GPDs \cite{Ji:1996ek}
by adding up the sum rules (\ref{Eq:EMT-and-GPD-H}) and (\ref{Eq:EMT-and-GPD-E}).

The model expression for $(H^u+H^d+E^u+E^d)(x,\xi,t)$, which we shall refer to
as $E_M(x,\xi,t)$ for brevity, reads \cite{Ossmann:2004bp}
\ba
    E_M(x,\xi,t) &=&
        \frac{iM_N^2N_c}{2I}\;\frac{\epsilon^{3ab} \Delta^a}{\bDelta_\perp^2}\,
    \int\!\frac{\di z^0}{2\pi}\,e^{iz^0xM_N}\Biggl[
    \biggl\{\doublesum{m,{\rm occ}}{j,{\rm all}}e^{-iz^0E_m}
               -\doublesum{m,{\rm all}}{j,{\rm occ}}e^{-iz^0E_j}\biggr\}
    \frac{1}{E_m-E_j}\nonumber\\
 &&     \times\,\la m|\tau^b|j\ra\,
        \la j|\,(1+\gamma^0\gamma^3)\,\exp(-iz^0\hat{p}^3/2)\,
    \exp(i\bDelta\hat{\bf X})\,\exp(-iz^0\hat{p}^3/2)\,|m\ra\nonumber\\
 && + iz^0\sum\limits_{m,\rm occ}e^{-iz^0E_m}
    \la m|\tau^b(1+\gamma^0\gamma^3)\,\exp(-iz^0\hat{p}^3/2)\,
    \exp(i\bDelta\hat{\bf X})\,\exp(-iz^0\hat{p}^3/2)\,|m\ra
    \Biggr]
    \label{E_M-model}\ea
Integrating the $xE_M(x,\xi,t)$ over $x$ yields, after substituting $x\to y = xM_N$
and extending the limits of $y$-integration to the entire $y$-axis which is
justified in the large $N_c$-limit due to $M_N={\cal O}(N_c)$, the following result
\ba
     \int\limits_{-1}^1\di x \;x E_M(x,\xi,t) =
         \frac{i N_c}{2I}\;\frac{\epsilon^{3ab} \Delta^a}{\bDelta_\perp^2}
     \;\Biggl[\biggl\{
         \doublesum{m,{\rm occ}}{j,{\rm all}}E_m
    -\doublesum{m,{\rm all}}{j,{\rm occ}}E_j\biggr\}\frac{1}{E_m-E_j}
    \la m|\tau^b|j\ra\,\la j|\,(1+\gamma^0\gamma^3)\;
        \exp(i\bDelta\hat{\bf X})\,|m\ra && \nonumber\\
    +\frac12\biggl\{
         \doublesum{m,{\rm occ}}{j,{\rm all}}
    -\doublesum{m,{\rm all}}{j,{\rm occ}}\biggr\}\frac{1}{E_m-E_j}
    \la m|\tau^b|j\ra\,\la j|\,(1+\gamma^0\gamma^3)\,
        \{\hat{p}^3,\exp(i\bDelta\hat{\bf X})\}\,|m\ra\; &&\nonumber\\
    -
    \sum\limits_{m,\,{\rm occ}} \la m|\tau^b\,(1+\gamma^0\gamma^3)\,
    \exp(i\bDelta\hat{\bf X})\, |m\ra \Biggr]\;. &&
    \label{E_M-model-02}\ea
Consider the unitary transformation $G_5\equiv\gamma^2\gamma^5\tau^2$
in the notation of \cite{Bjorken+Drell} with the properties
$G_5\gamma^\mu G_5^{-1} = (\gamma^\mu)^T$ and $G_5\tau^a G_5^{-1} = -(\tau^a)^T$.
It transforms in coordinate space the Hamiltonian (\ref{Hamiltonian}) and its
eigenstates as $G_5 H G_5^{-1}=H^T$ and $G_5\Phi_n({\bf x})=\Phi_n^\ast({\bf x})$.
Making use of this transformation we obtain the identities
\ba
    \la j|\tau^b|m\ra &=& \la m|(-\tau^b)|j\ra \nonumber\\
    \la j|\,(1+\gamma^0\gamma^3)\;\exp(i\bDelta\hat{\bf X})|m\ra
&=& \la m|\,(1-\gamma^0\gamma^3)\;\exp(i\bDelta\hat{\bf X})|j\ra \nonumber\\
    \la j|\tau^b\,(1+\gamma^0\gamma^3)\,\exp(i\bDelta\hat{\bf X})\, |m\ra
&=& \la m|\tau^b\,(-1+\gamma^0\gamma^3)\,\exp(i\bDelta\hat{\bf X})\,|j\ra\nonumber\\
    \la j|\,(1+\gamma^0\gamma^3)\;\{\hat{p}^3,\exp(i\bDelta\hat{\bf X})\}|m\ra
&=& \la m|\,(1-\gamma^0\gamma^3)\;\{-\hat{p}^3,\exp(i\bDelta\hat{\bf X})\}|j\ra
\ea
one obtains
\be
        \int\limits_{-1}^1\di x \;x E_M(x,\xi,t) =
        \frac{i N_c}{2I}\;\frac{1}{\bDelta_\perp^2}
        \doublesum{m,{\rm occ}}{j,{\rm non}}F_{mj}
    \label{E_M-model-03a}\ee
with
\ba
    F_{mj} &=& \;\;\, \frac{\epsilon^{ab3} \Delta^a}{E_m-E_j}\;
    \la m|\tau^b|j\ra\,\la j|
    \biggl[\,\gamma^0\gamma^3(E_m+E_j)\exp(i\bDelta\hat{\bf X})
    +\{\hat{p}^3,\exp(i\bDelta\hat{\bf X})\}\biggr]|m\ra\nonumber\\
    &=& \frac13\,\frac{\epsilon^{abc} \Delta^a}{E_m-E_j}\;
    \la m|\tau^b|j\ra\,\la j|
    \biggl[\,\gamma^0\gamma^c(E_m+E_j)\exp(i\bDelta\hat{\bf X})
    +\{\hat{p}^c,\exp(i\bDelta\hat{\bf X})\}\biggr]|m\ra
        \label{E_M-model-03b}\;,
\ea
where in the second step we made use of the hedgehog symmetry. The final step
necessary to recognize that (\ref{E_M-model-03a},~\ref{E_M-model-03b}) coincide
with the expression (\ref{Eq:J-model}) is the following.
Notice that in the expression in Eqs.~(\ref{E_M-model-03a},~\ref{E_M-model-03b})
there is no more reference to the 3-direction which was picked out by arbitrarily
choosing the space-direction of light-cone vector $n^\mu$ along the 3-axis.
Therefore it is justified to identify $\bDelta_\perp^2=\frac23\bDelta^2 = -\frac23\,t$,
which finally leads us to the result
\be     2J(t)_{\rm GPD} \equiv \int\limits_{-1}^1\di x \;x E_M(x,\xi,t) \equiv
    2J(t)_{\rm EMT} \;.
\ee

\section{Chiral properties of the form factors}
\label{App:chrial-properties}

In this Appendix we study the chiral properties of the EMT form factors,
and start with $d_1(t)$ because here we face simpler expressions and
the method is more easily explained.
We start from the expression (\ref{App2-Eq:M2-d1-from-GPD}) for $d_1(t)$,
and rewrite its continuum contribution in terms of the Feynman propagator,
see App.~A of Ref.~\cite{Diakonov:1996sr}, and expand it in gradients of
the $U$-field. In leading order of this expansion we obtain
\be\label{App-chiral-00}
    d_1(t)_{\rm cont}
    = \frac{75f_\pi^2M_N}{4\,t}\;
        \int\!\di^3{\bf x}\;\,j_2(r\sqrt{-t})\,P_2(\cos\theta)
        \,{\rm tr_F}[\nabla^3 U][\nabla^3 U^\dag] + \dots \ee
where the dots denote terms containing three or more gradients of the $U$-field.
Taking $t\to 0$ in (\ref{App-chiral-00}), and using
$j_l(z) = \frac{z^l}{(2l+1)!!} + {\cal O}(z^{l+2})$ for $z\ll 1$,
we recover the result for $d_1^{\rm cont}$ in Eq.~(44) of
Ref.~\cite{Schweitzer:2002nm}.\footnote{
    \label{Footnote-d1-grad}
    In \cite{Schweitzer:2002nm} in the leading order in gradient expansion
    it was estimated $d_{1,\,\rm grad}^{\,\rm LO\; cont}\approx - 9.46$
    to be compared to $d_{1,\,\rm exact}^{\,\rm cont} = - 8.34$ which is
    the exact result obtained here in proper time regularization, i.e.\
    also in this case the gradient expansion provides a useful estimate
    for the continuum contribution of a quantity. Let us take here
    the opportunity to correct an error in \cite{Schweitzer:2002nm}.
    The level contribution to $d_1$ is not zero --- contrary to the claim in
    Eq.~(43) of \cite{Schweitzer:2002nm}. Instead, it is $d_1^{\rm lev} = 4.88$
    with the self-consistent proper-time profile.
    In fact, the level contribution to $d_1$ (and to the pressure) cannot be zero.
    It plays a crucial role in establishing the stability of the soliton, see
    Sec.~\ref{Sec-7:pressure+shear}.
    We stress that reliable results which satisfy the stability condition and
    all other requirements can only be obtained from evaluating numerically the
    exact model expressions, with the correct self-consistent profile,
    see Sec.~\ref{Sec-7:pressure+shear}.}

The result in Eq.~(\ref{App-chiral-00}) can be used to study the chiral properties
of $d_1(t)$. For that the leading large-$r$ (long distance) behaviour of the
integrand in (\ref{App-chiral-00}) plays the crucial role.
Therefore for this purpose it is legitimate to neglect both,
the discrete level contribution which has exponential fall-off at large $r$,
and higher order terms in the gradient expansion which have additional power
suppression with respect to the leading order term in (\ref{App-chiral-00}).

In Eq.~(\ref{App2-Eq:M2-d1-from-GPD}) we have chosen as starting point
$d_1(t)$ is given in terms of $s(r)$, see App.~\ref{App:FFs-from-EMT-and-GPD}.
Therefore --- after taking the flavour-trace ${\rm tr_F}$ in (\ref{App-chiral-00}),
integrating out the angular dependence, restoring the integral over the full
solid angle, and comparing to Eq.~(\ref{Eq:d1-from-s(r)-and-p(r)}) --- we read
off the expression for $s(r)$,
\ba
    d_1(t)_{\rm cont}
    &=& \frac{5M_N}{t}\;\int\!\di^3{\bf x}\;j_2(r\sqrt{-t})\,s(r)
    \;\;\;\mbox{with}\;\;\;
    s(r)=f_\pi^2\biggl(P^\prime(r)^2-\frac{\sin^2P(r)}{r^2}\biggr)+\dots
    \label{App-chiral-01}\ea
Making use of the long-distance behaviour of the profile (\ref{Eq:profile-at-large-r})
we obtain for the large-$r$ behaviour of $s(r)$ in the chiral limit the result quoted
in Eq.~(\ref{Eq:pressure-03}). The large-$r$ behaviour of $p(r)$ quoted there
follows from (\ref{Eq:p(r)+s(r)}).

In order to derive the leading non-analytic contributions to $d_1(t)$
we choose to work with the following analytic form of the soliton profile
\be\label{App-chiral-02-atan-profile}
    P(r)=-2\arctan\biggl[\frac{R^2}{r^2}\,(1+m_\pi r)\;\exp(-m_\pi r)\biggr]\;.
\ee
This profile was demonstrated to be a good approximation to the true self-consistent
profile \cite{Diakonov:1987ty}. Since it does not correspond to the true minimum
of the soliton energy, e.g., the  approximate result for the pressure obtained with
this profile does not satisfy the stability condition (\ref{Eq:stability}). However,
all that matters for our purposes is that it exhibits the correct chiral behaviour,
see (\ref{Eq:profile-at-large-r}). The ``soliton radius'' $R$ in
(\ref{App-chiral-02-atan-profile}) is connected to the constant $A$ in
(\ref{Eq:profile-at-large-r}) by $A=2R^2$.

Let us focus on the chiral expansion of $d_1(t)_{\rm cont}$ at
zero momentum transfer. We obtain
\ba
    d_1^{\rm cont}
    &=& - \frac{4 \pi}{3}\,f_\pi^2M_NR^3 \,G(m_\pi R)
    \;\;\;\mbox{with}\;\;\;
    G(a)=\int\limits_0^\infty\!\di z\;\,
    \frac{4z^6(3 + 6az + 7a^2z^2+4a^3z^3+a^4z^4)e^{-2az}}
         {(z^4+(1+az)^2e^{-2az})^2}
    \label{App-chiral-03}\ea
The zeroth order in the Taylor series of the function $G(a)$ around $a=0$ is
$G(0) = \int_0^\infty\di z\,12 \,z^6/(z^4+1)^2 = 9\pi/\sqrt{8}$.

To find the linear term in the Taylor expansion of $G(a)$ we proceed as follows,
c.f.\  App.~B of \cite{Schweitzer:2003sb} for a similar calculation.
We consider $G^\prime(a)$ at $a\neq 0$, substitute $z\to y=az$,
and consider then the limit $a\to0$.
  Notice that, had we decided to evaluate (\ref{App-chiral-03}) in a finite volume,
  let us say in a spherical box of radius $D$,
  then the upper limit of the $y$-integration would be $Dm_\pi$.
  Thus, before taking the chiral limit it is crucial to take first
  the infinite volume limit $D\to\infty$ \cite{Schweitzer:2003sb},
  as these two limits do not commute.
This yields
$G^\prime(a)|_{a=0} = -8\int_0^\infty\di y\;(-1+y+2y^2+y^3)\,e^{-2y}=-5$.
Thus, we obtain
\be\label{App-chiral-06}
    G(a) = G(0) - 5 a + \mbox{higher order terms in a}.
\ee
Inserting this result in (\ref{App-chiral-03}) we reproduce the chiral limit result
in Eq.~(46) of \cite{Schweitzer:2002nm} which, however, provides just one contribution
to the chiral limit value of $d_1$. Other contributions are of importance,
see Footnote~\ref{Footnote-d1-grad}.

The situation is different for the linear $m_\pi$-correction to $d_1$.
As explained above, here the expansion (\ref{App-chiral-06}) provides, in fact,
the correct leading non-analytic term in the chiral expansion of the full
expression for $d_1$ in the model. Combining  (\ref{App-chiral-03}) and
(\ref{App-chiral-06}) and eliminating $R \equiv \sqrt{A/2}$ in favour of
$g_A$ and $f_\pi$ according to (\ref{Eq:profile-at-large-r}) we obtain
Eq.~(\ref{Eq:d1-vs-mpi}). The result for $d_1^\prime(0)$
in (\ref{Eq:d1-slope-at-t=0-chir-lim}) is obtained similarly.

To derive the small $t$ expansion of $d_1(t)$ in the chiral limit
in Eq.~(\ref{Eq:d1-at-small-t-chir-lim}) one can integrate (\ref{App-chiral-01})
exactly. The result is a bulky and not illuminating expression which we do not
quote here. Expanding it for small $t$ yields (\ref{Eq:d1-at-small-t-chir-lim}).

Let us turn to the discussion of the form factor $M_2(t)$. At $t=0$ we have
for all $m_\pi$ exactly $M_2(0)=1$, see Eq.~(\ref{App-constraints-01}).
The slope of $M_2(t)$ at zero momentum transfer, however, has a non-trivial
chiral expansion.
Due to (\ref{Eq:M2-slope-at-t=0}) we need for that the chiral expansion of
$d_1(t)$, see above, and that of the mean square radius of the energy density.
In the above described way we obtain from the expression
(\ref{Eq:energy-density-in-grad-expansion}) for the energy density
in the leading order gradient expansion the following contribution
\be\label{App-chiral-07}
    \la r_E^2\ra_{\rm part\,1} = \la \krig{r}{\!}_E^2\ra
    -\frac{23}{2}\;(4\pi f_\pi^2 R^4)\;\frac{m_\pi}{M_N }+\dots
\ee
which is, however, not yet the complete result for the following reason.
The chiral expansion of the exact model expression for the energy density
(\ref{Eq:density-energy}) in the gradient expansion contains --- in addition
to Eq.~(\ref{Eq:energy-density-in-grad-expansion}) --- also the term
\be\label{App-chiral-08}
    \rho_E(r)_{\rm part\,2} = \frac{m_\pi^2 f_\pi^2}{4} \;{\rm tr}_F(2-U-U^\dag)
    = m_\pi^2 f_\pi^2 \biggl(1-\cos P(r)\biggr)\;.
\ee
This term is of ``zeroth order'' in the gradient expansion. It arises
from the current quark mass term in (\ref{eff-theory}) and is related to the
nucleon-pion sigma-term in the gradient expansion \cite{Schweitzer:2003sb}.
The explicit appearance of the current quark mass has been eliminated
in (\ref{App-chiral-08}) in favour of $m_\pi^2$ by means of the
Gell-Mann--Oakes--Renner relation which is valid in the model.
The $m_\pi^2$-term (\ref{App-chiral-08}) vanishes in the chiral limit,
and was therefore not displayed in (\ref{Eq:energy-density-in-grad-expansion}),
but it contributes in linear $m_\pi$-order to (\ref{App-chiral-07}).
This may not be obvious at a first glance, however, from (\ref{App-chiral-08})
we obtain
\be\label{App-chiral-09}
    \la r_E^2\ra_{\rm part\;2} =
    -\,4\pi f_\pi^2 R^4\;\frac{m_\pi}{M_N }\int\limits_0^\infty\di z\;
    \frac{z^8(z^3-2z-1)e^{-2z}-4a^4z^4(1+z)^4}{(z^4+a^4(1+z)^2e^{-4z})^2}
    \Biggr|_{a= 0}\;+\dots
    = \frac{5}{2}\;(4\pi f_\pi^2 R^4)\;\frac{m_\pi}{M_N }+\dots
    \;,
\ee
and adding up (\ref{App-chiral-07}) and (\ref{App-chiral-09}) we obtain
\be\label{App-chiral-10}
    \la r_E^2\ra = \la \krig{r}{\!}_E^2\ra
    -9\;(4\pi f_\pi^2 R^4)\;\frac{m_\pi}{M_N }+\dots
\ee
It is not necessary to consider chiral corrections due to the nucleon mass
$M_N=\krig{M}_N+Bm_\pi^2+\dots$ because they contribute only at higher orders.
(Notice that the CQSM also consistently describes the chiral expansion of $M_N$
\cite{Schweitzer:2003sb}.)
Eliminating $R=\sqrt{A/2}$ in (\ref{App-chiral-10}) by means of
(\ref{Eq:profile-at-large-r}) yields finally the results for $\la r_E^2\ra$
in (\ref{Eq:rE2-vs-mpi}) and for $M_2^\prime(0)$ in (\ref{Eq:M2-slope-at-t=0}).

Finally we comment on the form factor $J(t)$. Also in this case the normalization
is trivial, since $J(0)=\frac12$, but e.g.\ the chiral expansion of
$J^\prime(0)=\frac16\,\la r_J^2\ra$ is of interest. Here we restrict ourselves
to the mere observation that $\la r_J^2\ra$ diverges in the chiral limit.

In contrast to $M_2(t)$ and $d_1(t)$, which are non-zero in the leading order of
the large-$N_c$ expansion, $J(t)$ arises from $1/N_c$ (``rotational'') corrections.
For such quantities the non-commutativity of the large-$N_c$ and chiral limit
may have more drastic implications \cite{Cohen:1992uy}.
For example, in the slowly rotating soliton approach
(as realized e.g.\ in the Skyrme model in Ref.~\cite{Adkins:1983hy})
the isovector electric mean square radius diverges as $1/m_\pi$ in the
chiral limit --- in contrast to ${\rm ln }\,m_\pi$ at finite $N_c$.
For $\la r_J^2\ra$ the situation is completely analog --- as a study
in the Skyrme model reveals \cite{Skyrme-preprint}.

\newpage

\end{document}